\documentclass[acmlarge]{acmart}
\AtBeginDocument{%
  \providecommand\BibTeX{{%
    \normalfont B\kern-0.5em{\scshape i\kern-0.25em b}\kern-0.8em\TeX}}}

\setcopyright{acmlicensed}
\acmJournal{IMWUT}
\acmYear{2024} \acmVolume{8} \acmNumber{4} \acmArticle{189} \acmMonth{12}\acmDOI{10.1145/3699741}
\begin{teaserfigure}
  \centering
  \includegraphics[width=\columnwidth]{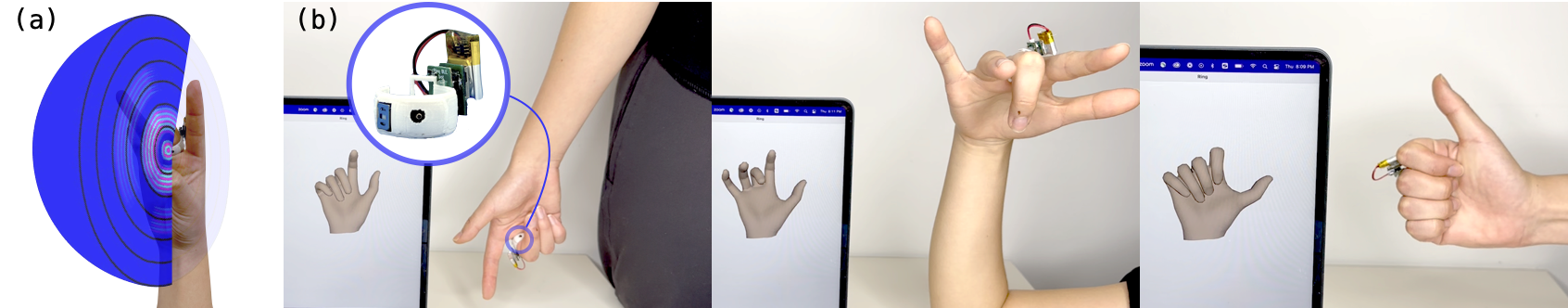}
  \caption{~\theDevice{} is an untethered ring that tracks hand poses continuously. With active acoustic sensing, we analyze the reflection strengths at different distances from the ring to reconstruct hand poses. (a) shows an example reflection profile. (b) shows examples of ~\theDevice{}'s real-time inferences at different wrist and forearm orientations.
  }\label{fig: teaser}
\end{teaserfigure}

\usepackage{color,soul}
\usepackage{booktabs} 
\usepackage{subfig} 
\usepackage{graphicx}

\usepackage[utf8]{inputenc}
\usepackage[T1]{fontenc}

\usepackage{siunitx}
\newcommand{\hide}[1]{}

\sethlcolor{yellow}
\ifdefined\scifihidecomments
    \newcommand{\cz}[1] {}
    \newcommand{\hyunc}[1] {} 
    \newcommand{\yax}[1] {} 
    \newcommand{\sony}[1] {} 
    \newcommand{\ReviewerFeedback}[1] {}  
    \newcommand{\fran}[1] {}
    \newcommand{\rz}[1]{}
    \newcommand{\lw}[1] {}
    \newcommand{\mose}[1] {} 
    \newcommand{\ke}[1] {} 

\else
    \definecolor{burntorange}{rgb}{0.8, 0.33, 0.0}
    \definecolor{cadmiumgreen}{rgb}{0.0, 0.42, 0.24}
    \definecolor{cobalt}{rgb}{0.0, 0.28, 0.67}
    \definecolor{amber}{rgb}{1.0, 0.75, 0.0}
    \definecolor{fashionfuchsia}{rgb}{0.96, 0.0, 0.63}
    \definecolor{brightcerulean}{rgb}{0.11, 0.67, 0.84}
    \definecolor{frenchblue}{rgb}{0.0, 0.45, 0.73}
    \definecolor{darkslateblue}{rgb}{0.28, 0.24, 0.55}
    \definecolor{cerulean}{rgb}{0.0, 0.48, 0.65}
    \definecolor{darkpastelgreen}{rgb}{0.01, 0.75, 0.24}

    \newcommand{\cz}[1] { \textcolor{red}{[\hl{cheng:} {#1}}]}
    \newcommand{\hyunc}[1] { \textcolor{burntorange}{[{\hl{hyunc:}} {#1}}]}
    \newcommand{\yax}[1] { \textcolor{magenta}{[{\hl{yaxuan:}} {#1}]}}
    
    \newcommand{\sony}[1] { \textcolor{blue}{[{\hl{songyun:}} {#1}]}}
    \newcommand{\ReviewerFeedback}[1] { \textcolor{brightcerulean}{[{Reviewer Feedback:} {#1}}]}
    \newcommand{\fran}[1]{\textcolor{burntorange}{[{francois:}{#1}}]}
    \newcommand{\rz}[1]{\textcolor{teal}{[{Ruidong: }{#1}]}}
    \newcommand{\lw}[1]{\textcolor{fashionfuchsia}{[{liuwei:}{#1}}]}
    \newcommand{\mose}[1]{\textcolor{burntorange}{[{mose:}{#1}}]}
    \newcommand{\ke}[1] { \textcolor{red!55!yellow}{[{Ke:} {#1}}]}

\definecolor{CAT-comment}{rgb}{0.95, 0.2, 0.8}
\definecolor{GL-comment}{rgb}{0.0, 0.54, 0.8}
\definecolor{RR-changes}{rgb}{0.0, 0.26, 0.91}

\newcommand{\etal}{et al.~}
\fi

\ifdefined\scifiblind
    \newcommand{\blind}[1]{[omitted for blind review]}
\else
    \newcommand{\blind}[1]{#1} 
\fi

\newcommand{\theDevice}{Ring-a-Pose}
\begin{document}

\title{\theDevice: A Ring for Continuous Hand Pose Tracking}

\author{Tianhong Catherine Yu}
\email{ty274@cornell.edu}
\orcid{0000-0002-3742-0178}
\affiliation{
  \institution{Cornell University}
  \city{Ithaca}
  \state{New York}
  \country{USA}
}

\author{Guilin Hu}
\email{gh386@cornell.edu}
\orcid{0009-0001-6404-9968}
\affiliation{
  \institution{Cornell University}
  \city{Ithaca}
  \state{New York}
  \country{USA}
}

\author{Ruidong Zhang}
\email{rz379@cornell.edu}
\orcid{0000-0001-8329-0522}
\affiliation{
  \institution{Cornell University}
  \city{Ithaca}
  \state{New York}
  \country{USA}
}

\author{Hyunchul Lim}
\email{hl2365@cornell.edu}
\orcid{0000-0001-8397-3534}
\affiliation{
  \institution{Cornell University}
  \city{Ithaca}
  \state{New York}
  \country{USA}
}

\author{Saif Mahmud}
\email{sm2446@cornell.edu}
\orcid{0000-0002-5283-0765}
\affiliation{
  \institution{Cornell University}
  \city{Ithaca}
  \state{New York}
  \country{USA}
}

\author{Chi-Jung Lee}
\email{cl2358@cornell.edu}
\orcid{0000-0002-1887-4000}
\affiliation{
  \institution{Cornell University}
  \city{Ithaca}
  \state{New York}
  \country{USA}
}

\author{Ke Li}
\email{kl975@cornell.edu}
\orcid{0000-0002-4208-7904}
\affiliation{
  \institution{Cornell University}
  \city{Ithaca}
  \state{New York}
  \country{USA}
}

\author{Devansh Agarwal}
\email{da398@cornell.edu}
\orcid{0009-0005-1338-9275}
\affiliation{
  \institution{Cornell University}
  \city{Ithaca}
  \state{New York}
  \country{USA}
}

\author{Shuyang Nie}
\email{sn429@cornell.edu}
\orcid{0009-0004-1375-4626}
\affiliation{
  \institution{Cornell University}
  \city{Ithaca}
  \state{New York}
  \country{USA}
}

\author{Jinseok Oh}
\email{jo293@cornell.edu}
\orcid{0009-0004-7095-6910}
\affiliation{
  \institution{Cornell University}
  \city{Ithaca}
  \state{New York}
  \country{USA}
}

\author{François Guimbretière}
\email{fvg3@cornell.edu}
\orcid{0000-0002-5510-6799}
\affiliation{%
  \institution{Cornell University}
  \city{Ithaca}
  \state{New York}
  \country{USA}
}

\author{Cheng Zhang}
\email{chengzhang@cornell.edu}
\orcid{0000-0002-5079-5927}
\affiliation{%
  \institution{Cornell University}
  \city{Ithaca}
  \state{New York}
  \country{USA}
}
\renewcommand{\shortauthors}{Yu et al.}

\begin{abstract}
We present Ring-a-Pose, a single untethered ring that tracks continuous 3D hand poses. Located in the center of the hand, the ring emits an inaudible acoustic signal that each hand pose reflects differently. Ring-a-Pose imposes minimal obtrusions on the hand, unlike multi-ring or glove systems. It is not affected by the choice of clothing that may cover wrist-worn systems. In a series of three user studies with a total of 36 participants, we evaluate Ring-a-Pose's performance on pose tracking and micro-finger gesture recognition. Without collecting any training data from a user, Ring-a-Pose tracks continuous hand poses with a joint error of 14.1mm. The joint error decreases to 10.3mm for fine-tuned user-dependent models. Ring-a-Pose recognizes 7-class micro-gestures with a 90.60\% and 99.27\% accuracy for user-independent and user-dependent models, respectively. Furthermore, the ring exhibits promising performance when worn on any finger. Ring-a-Pose enables the future of smart rings to track and recognize hand poses using relatively low-power acoustic sensing.

\end{abstract}






\maketitle

\section{Introduction}
The smart ring is a promising and rising wearable for everyday use~\cite{OuraRing}.
However, due to the spatial constraint, the sensing capability of the ring has been limited: commodity rings integrate miniaturized and low-power sensors (\textit{e.g.}, heart rate sensors, IMUs).
Despite being worn on the finger, the existing smart ring, on its own, can not track hand poses continuously, which is crucial in understanding the user's actions and intentions (e.g., VR/AR interaction, human activity recognition).
To address this problem, prior work has explored different sensing modalities to recognize hand/finger gestures on a ring~\cite{Cyclopsring, magicring, iRing, energyinputring, thumbtrak, z-ring, FingerPing, zhang2017fingersound, zhang2017fingorbits}, but multiple rings along with a wristband ~\cite{learning-on-the-rings, onering} are required to track full hand poses, not desirable for everyday uses.
Continuous hand pose tracking is particularly complex as hand joint movements have 21 degrees of freedom involving finger flexions/extensions and abductions/adductions ~\cite{learning-on-the-rings}.
The key challenge with a ring is to sense enough information on the pose and movement of each finger from a single ring-worn position, which is challenging to capture using traditional motion sensors like IMUs.
Z-Pose explores bio-impedance measurements from a ring for hand pose tracking, but generalizing across sessions and users proves difficult due to inconsistent impedance measurements~\cite{z-pose}.
In this paper, we seek to answer the research question of \textbf{whether we can continuously track hand poses across sessions and users with just a single untethered ring}.

To this end, we present~\theDevice{} (Fig.~\ref{fig: teaser}), a single untethered ring that tracks full hand poses continuously using active acoustic sensing.
It only requires a microphone and a speaker embedded into the ring, which already (partially) exist in commodity smart rings (\textit{e.g.},~\cite{OuraRing}). 
Facing inside the palm, the speaker emits inaudible frequency-modulated continuous waves (FMCW) around the fingers, which are reflected by the palm and fingers with unique patterns and received by the co-located microphone.
Using the correlation-based frequency modulated continuous wave method (C-FMCW)~\cite{C-FMCW}, we obtain the reflection strengths at different echo path lengths from the ring.
In addition to capturing the movement of the instrumented finger like IMUs, our ring sensor also captures the movements of other fingers as their movements alter the reflection even when not directly instrumented.
~\theDevice{} is low-cost ($\sim$US\$30) with a low profile. Furthermore, the battery consumption is relatively low power (148.0 mW) with the possibility of further optimization.

Active acoustic sensing on wearables exhibited exciting potential for capturing fine-grained body part movements from echos~\cite{EarIO, EchoWrist, Eyeecho}, but the ring form factor and worn location imposes additional sensing challenges: (a) \textit{limited physical space}: to preserve the ring's low-profile nature, we embed only one pair of microphone and speaker into the ring (differing from multiple pairs of speakers and microphones from prior works), at the cost of reduced sensed information for a complex reconstruction task; and (b) \textit{sensor movement\&occlusion caused by finger movements}: the ring form factor has the unique benefit of being "inside" the hand that avoids the orientation disparity between the wrist and the sensor suffered by wrist-mounted devices~\cite{EchoWrist,discoband,etherpose}, but the sensor moves along with the worn finger and can be occluded by the fingers.
Embracing the challenges in pursuit of a compact yet powerful ring, we develop a lightweight deep-learning pipeline.
The unique sensor placement, inside the hand, allows the ring to (a) capture rich information about the entire hand with just one input channel; and (b) turn finger occlusions into repeatable information for the reconstruction task.
We further discuss this in Sec.~\ref{sec:echo-profiles-for-ring}.

To thoroughly evaluate the performance of~\theDevice{}, we conducted three studies. 
In the first study with 10 participants, we compared the performance when the ring was worn on different fingers. The results showed that ~\theDevice{} can track the hand poses well on any worn finger, which was underexplored in prior ring-based sensing systems~\cite{RingReview}. 
In the second study with 12 participants, we showed that ~\theDevice{} achieves a mean per-joint position error (MPJPE) of 14.1 mm and 10.3 mm in user-independent and user-dependent evaluations, respectively. 
In the third user study with 10 participants, ~\theDevice{} was able to recognize 7 micro-finger gestures with accuracies of 90.60\% and 99.27\% in user-independent and user-dependent evaluation respectively. 

Unlike most data-driven hand pose tracking systems that have significant performance degradation when the sensor is remounted, ~\theDevice{} is independent of wearing sessions.
Additionally, ~\theDevice{} exhibits exciting performance even when used out-of-the-box, i.e. no calibration data from the new user, further highlighting the practicality of ring-based hand pose tracking systems. 
In summary, the main contributions of this paper are:
\begin{itemize}
    \item To the best of our knowledge, ~\theDevice{} is the first single-ring system to demonstrate continuous hand pose tracking across sessions and participants: we applied the C-FMCW method to the ring to reconstruct hand poses based on the reflection strengths at different echo path lengths from the ring.
    \item We validated that the ring effectively tracks the hand pose when worn on any finger with finger-dependent models.
    \item We evaluated ~\theDevice{}'s performance in continuous hand pose tracking and micro-finger gesture recognition with user studies.
    \item We further discuss the opportunities and challenges of integrating ~\theDevice{} to the future commodity ring platforms. 
\end{itemize}

\section{Related Work}
Motivated by attractive applications of hand pose tracking in natural interaction, rehabilitation, immersive AR/VR experiences, sign language translation, etc., the computer vision community has explored various robust optical sensing approaches off-the-body: multi-camera systems~\cite{Vicon}, monocular RGB cameras \cite{mueller2018ganerated, mediapipe}, and depth cameras \cite{leapmotion, Qian2014}.
However, sensor instrumentation in the environment lacks portability and the occlusion issue limits the users' movements.
Thus, there is a need for reliable wearable approaches that are portable and suffer less from occlusions from the environment.
Here, we focus on reviewing wearable approaches that are pertinent to our contribution.
In the rest of the section, we divide related works into three key areas: (1) hand pose sensing, the functionality of~\theDevice{}, (2) acoustic sensing, the sensing technique of~\theDevice{}, and (3) sensing rings, the form factor of~\theDevice{}.

\subsection{Hand Pose Sensing on Wearables}\label{sec: rw-handpose}
Mounting the sensor directly on the user allows mobile hand sensing.
Cameras mounted on the heads are suitable for VR uses~\cite{VisionPro,quest}, but are not practical for day-to-day uses.
Wrist is a popular instrumentation site.
Wrist-worn devices have low proximity to the hand and do not fully cover the hand like data gloves~\cite{stretch-glove, imu-glove-clinic} which are unrealistic to be worn at all times.
Cameras mounted on the wrists~\cite{fingertrak, Back-hand-pose, opisthenar} and other range-finding sensors~\cite{discoband, thumbtrak, beamband, EchoWrist} can reconstruct hand poses or recognize hand gestures from limited viewpoints, but they require a clear line-of-sight from the wrist that cannot be covered by long-sleeve clothing.
Other hand pose tracking sensing principles that do not suffer from sensor occlusion include impedance characteristics~\cite{etherpose, z-pose} that do not generalize across users and electromyography (EMG)~\cite{NeuroPose} with a bulky form factor.
Our ring is not affected by clothing choices, generalizes well across users, and is low-profile.

Hand gesture recognition is an easier sensing task than hand pose tracking but is crucial to gestural controls.
Prior systems of full-hand gesture recognition are implemented with impedance~\cite{tomo}, wrist pressure~\cite{DementyevPressure2014}, ultrasonic beamforming~\cite{beamband}, capacitive sensing~\cite{capband}, etc.
To enable more discrete and acceptable natural interactive contorl~\cite{ConsensusGesture}, micro-finger gesture recognition that requires fine-grained sensing is gaining traction ~\cite{sparseimu, solimicro, Thumb-In-Motion,pyro, handsense, zhang2017fingorbits, FingerPing}.
In our work, we demonstrate that ~\theDevice{} effectively recognizes micro-finger gestures.

\subsection{Acoustic Sensing on Wearables}\label{sec: rw-acoustic}
On-body acoustic sensing has emerged as a reliable approach for tracking contexts and movements due to its robustness to noise factors like lighting conditions and electric fields.
The applications span 
hand gesture sensing~\cite{beamband, FingerPing, Skinput, EchoWrist}, motion tracking~\cite{wang2019millisonic, MaoCat16, Earphonetrack}, pose tracking\cite{PoseSonic}, activity recognition~\cite{Bodyscope}, food intake recognition~\cite{AutoDietary}, facial expression tracking~\cite{EarIO, Eyeecho}, teeth and tongue gesture recognition\cite{echospeech,sun2021teethtap}, gaze tracking~\cite{li2024gazetrak}, silent speech recognition~\cite{echospeech,JinEarCommand22,zhang2023hpspeech,sun2023echonose}, etc. 

The most recent work, EchoWrist~\cite{EchoWrist} is the closest to our system. It used two pairs of microphones and speakers on a wristband to track hand poses and recognize hand-object interactions. 
For a similar sensing task, designing a sensing system for a ring exhibits much greater challenges than for a wristband as the ring has a smaller physical space. 
For example, a ring can not afford two pairs of microphones and speakers due to spatial constraints and energy consumption considerations. 
Although both systems employ acoustic sensing, EchoWrist and~\theDevice{} adapt different sensing principles to infer hand poses: EchoWrist infers hand poses and hand-object interaction from the contour shape around the wrist captured by acoustic sensors, while Ring-a-Pose infers hand poses and recognizes hand gestures from the acoustic reflection from the fingers directly. 
The ring benefits from its unique worn location. 
Long-sleeved clothing easily covers the wristband but not the ring.
The ring's closer proximity to the fingers allows stronger acoustic reflections from the fingers and enables (a) fine-grained tracking like micro-finger gestures, which is hard to capture information from the wristband\cite{EchoWrist} and (b) similar tracking performance with only half the number of the speaker and the microphone. 
Last, unlike EchoWrist, ~\theDevice{} does not necessitate returning to a neutral hand pose between transitioning between poses (further detailed in Sec.~\ref{sec:pose-transitions}). We are the first to explore active acoustic sensing on the form factor of a ring to enable accurate continuous hand pose tracking and micro-finger gesture recognition.

\begin{table}[t]
\footnotesize
\caption{A High-Level Overview of Hand Sensing Rings.~\theDevice{} is the first to track continuous hand poses with just a single ring. In the labels, ``UI'' stands for user-independence and ``UD'' stands for user-dependence.}
\label{table:rings}
\begin{tabular}{|c|c|c|c|c|c|c|c|}
\hline
\textbf{System} & \textbf{\begin{tabular}[c]{@{}c@{}}Tracking \\ Output\end{tabular}} & \textbf{Sensor} & \textbf{\begin{tabular}[c]{@{}c@{}}Power \\ Consumption\end{tabular}} & \textbf{\begin{tabular}[c]{@{}c@{}}\# of \\ Components\end{tabular}} & \textbf{\begin{tabular}[c]{@{}c@{}}Un-\\ tethered\end{tabular}} & \textbf{UI} & \textbf{UD} \\ \hline
ElectroRing, 2021 \cite{ElectroRing21} & Discrete Tap & {\color[HTML]{333333} Electric Field} & 220mW & 1 Ring & \checkmark & {\color[HTML]{333333} $\times$} & \checkmark \\ \hline
ThumbTrak, 2018 \cite{thumbtrak} & Discrete Gesture & Proximity & 120mW & 1 Ring & $\times$ & $\times$ & \checkmark \\ \hline
FingerPing, 2018 \cite{FingerPing} & Discrete Gesture & Active Acoustic & - & 1 Ring, Wristband & $\times$ & $\times$ & \checkmark \\ \hline
Boldu et al., 2018 \cite{Thumb-In-Motion} & Discrete Gesture & Capacitive & 475mW & 1 Ring, Wristband & \checkmark & \checkmark & \checkmark \\ \hline
CyclopsRing, 2015 \cite{Cyclopsring} & Discrete Gesture & Camera & - & 1 Ring & $\times$ & \checkmark & \checkmark \\ \hline
EFRing, 2022 \cite{efring} & \begin{tabular}[c]{@{}c@{}}Discrete Gesture\\ 1D Continuous\end{tabular} & Electric Field & - & 1 Ring & $\times$ & \checkmark & \checkmark \\ \hline
Z-Ring, 2023 \cite{z-ring} & \begin{tabular}[c]{@{}c@{}}Discrete Gesture\\ 2D Continuous\end{tabular} & Bio-Impedance & 2.4W & 1 Ring, Wristband & $\times$ & \checkmark & \checkmark \\ \hline
Zhou et al., 2023 \cite{onering} & Continuous Pose & IMU and PPG & 44mW/ring & 2-5 Rings, Wristband & \checkmark & \checkmark & \checkmark \\ \hline
ssLOTR, 2022 \cite{learning-on-the-rings} & Continuous Pose & IMU & 198mW & 2-5 Rings, Wristband & {\color[HTML]{333333} \checkmark} & \checkmark & \checkmark \\ \hline
\textbf{Ring-a-Pose} & \textbf{Continuous Pose} & \textbf{Active Acoustic} & \textbf{148mW} & \textbf{1 Ring} & \checkmark & \checkmark & \checkmark \\ \hline
\end{tabular}
\end{table}
\subsection{Sensing Rings}\label{sec: rw-ring}
Commodity smart rings today specialize in fitness\&wellness~\cite{OuraRing,ringly} and contactless payment~\cite{RingPay}, missing the hand-related interaction space unveiled uniquely by the ring form factor.
Researchers tried to fill the gap with IMUs~\cite{vatavu2023ifad, gheran2018gestures, learning-on-the-rings}, proximity sensors ~\cite{thumbtrak}, electric field sensing ~\cite{z-ring,efring,TelemetRing}, capacitive sensing~\cite{ElectroRing21,Thumb-In-Motion}, electromagnetic sensing \cite{auraring}, infrared sensor~\cite{iRing}, minature cameras\cite{Cyclopsring, EyeRing}, acoustic signals \cite{ZhangRing11, energyinputring}, etc.
Sensing rings applications span text input \cite{nirjon2015typingring, gu2020qwertyring,TelemetRing}, health sensing \cite{poongodi2022diagnosis, onering}, authentication~\cite{z-ring, vibering}, and gestural inputs~\cite{ringdrone, FingerPad,auraring,Cyclopsring,RingReview}.
Vatavu and Bilius~\cite{RingReview} reviewed hand gesture inputs with rings, ring-like, and ring-ready devices.
\theDevice{} falls in the ``rings'' category: finger-worn device with a ring form factor.
In past literature, only multi-ring systems~\cite{learning-on-the-rings, onering} can track hand pose continuously.
ssLLOR~\cite{learning-on-the-rings} tracks hand poses with a wristband and 2-5 rings, each embedded with an IMU unit.
As shown in Table~\ref{table:rings}, ~\theDevice{} presents the first untethered ring that tracks 20 DoF hand poses with just a single ring across sessions and users, extending practical everyday smart ring capabilities.
Note the sensing principle is not the only factor determining energy consumption.
Other factors like microcontroller and communication method choices make the direct comparison unfair, but we include the overall power signature to better situate our relatively low-power system. 
One benefit of our active acoustic sensing approach is that the ring senses all finger movements well (unlike IMUs), no matter which finger the ring is instrumented on.
Thumb-In-Motion~\cite{Thumb-In-Motion} employs capacitive sensing on an index finger ring to sense thumb-to-index micro gestures.
~\theDevice{} recognizes thumb-to-index micro gestures with the ring on the middle finger and demonstrates the system's potential to detect fine-grain movements even when the moving fingers are not instrumented.
Furthermore, it is important to acknowledge that the ring has a limited space for electronics (especially the battery), which makes it challenging to prototype an untethered device.

\section{\theDevice{} Implementation}
The sensing ring, a novel form factor for continuous hand pose tracking, places the sensors on the finger, inside the hand.
This affords unique opportunities to simplify the hand reconstruction task: maintaining great visibility of the hand and consistency between the orientations of the wrist and the sensor.

\subsection{Design Objective}
Our design aims to preserve the slim and miniaturized form factor of a ring for comfortable prolonged wear while achieving promising sensing performance.
The complex sensing task that predicts 3D coordinates of 20 hand joints requires information about the thumb and all other 4 fingers. 
Thus, working with the space constraint, we experiment with only one speaker and one microphone instead of multiple speakers and microphones~\cite{echospeech, EarIO, beamband, EchoWrist}. 
Furthermore, to minimize the width of the ring, the speaker and microphone are placed side by side horizontally, instead of vertically.

\subsection{Form Factor Design}
We started the prototype process with rigid PCB islands but we moved to the flexible printed circuit board (FPCB) for a slimmer ring body and a better fit to the ring curvature.
The speaker and the microphone should be as close to each other as possible for accurate round-trip propagation time~\cite{C-FMCW}, but due to the ring curvature and thickness of the components, they are placed 5.4mm away from each other.
Our ring has a width of 11 mm.
Excluding the battery and the microcontroller (MCU), the ring body's thickest part, containing the speaker and microphone, is 3.58 mm thick, and the rest is 2 mm thick.
Fig.~\ref{fig: hardware}(c) shows a physical mockup of future~\theDevice{} with replaced arc battery and more minuturized MCU, comparable with that of the commercial Oura Ring~\cite{OuraRing}.
The ring is also lightweight, weighing 4.3 grams including the battery (1.8g).
Finally, to fit the finger sizes of all user study participants, we opted for a 3/4 circle ring and used yarn to adjust the ring size with a sliding knot.

\subsection{Hardware}
\begin{figure}[h]
  \includegraphics[width=\textwidth]{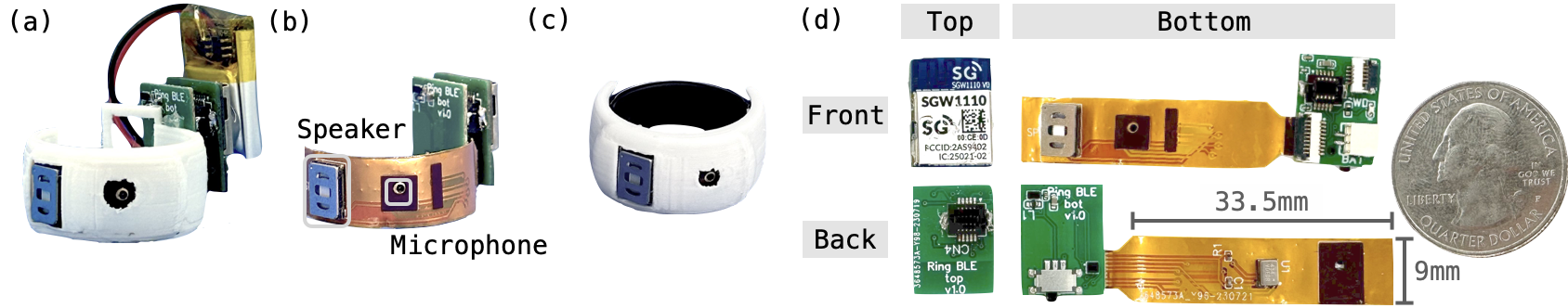}
  \caption{(a) The untethered ~\theDevice{} prototype used in the user studies. (b) The ~\theDevice{} prototype electronics without the case. (c) A physical mockup of future~\theDevice{}. We replaced the battery with an arc battery and removed the MCU. (d) Details of ~\theDevice{}'s PCBs.}
  \label{fig: hardware}
\end{figure}
The ring includes (a) a ring body, a customized FPCB enclosed in a 3D-printed PLA case, and (b) a customized capturing and processing PCB.
Fig.~\ref{fig: hardware}(b) shows the assembled hardware.
The top PCB (left in Fig.~\ref{fig: hardware}(d)) houses the SGW1110 module, featuring the nRF52840 microcontroller unit (MCU).
The low-power MCU implements Bluetooth 5 and provides Bluetooth Low Energy (BLE) functions with a built-in PCB-mounted antenna.
The bottom PCB, right in Fig.~\ref{fig: hardware}(d), contains an audio amplifier (MAX98357A), a voltage regulator (TPS62743) that provides a constant 3.3V source, a switch that turns ON/OFF the ring, a Flexible Printed Circuit (FPC) connector that connects to the ring body, and a battery connector.
The ring body FPCB features a speaker (USound UT-P2019) and a microphone (TDK 
ICS-43434).
The total cost of the prototype is about US\$30 and could be decreased when mass manufactured.
With the 3.7V 70mAh Lipo Battery, we measured (using a CurrentRanger) the energy consumption of the ring to be 148.0 mW.
The energy consumption breaks down into the MCU consuming 24.0 mW, the speaker and the microphone consuming 120.0 mW, and the BLE transmission consuming 4.0 mW on average.
We will further discuss the energy consumption in Sec.~\ref{sec: discussion}.
Similar to ~\cite{EarIO}, the speaker emits FMCW signals in the range of 20-24KHz with 600 samples per period, outside of the commonly stated range of human hearing, and the microphone samples at 50kHz.
The 16-bit sampled signals are transmitted to a nearby device (smartphone or computer) using BLE UART communication at 800 Kbps.

\subsection{Sensing Principle}\label{sec: sensing-principle}
We use active acoustic sensing as the sensing method for the ring to track continuous hand poses.
The ring emits inaudible sound waves, which are then reflected and refracted by the surrounding fingers and the palm.
As a result, the sound waves are received by the microphone with unique patterns.
As shown in Fig.~\ref{fig: echo-gests}, different hand geometries yield distinct reflected signals (i.e., echos) which we later process to reconstruct hand poses.

\subsubsection{Calculating Echo Profile using C-FMCW}
Wang~\etal{}proposed the correlation-based frequency modulated continuous wave method (C-FMCW) which on commodity audio devices, achieves higher ranging resolution, than that of traditional FMCW methods (34.3mm upper bound) ~\cite{C-FMCW}.
For the C-FMCW method, the theoretical ranging resolution, $R_{C-FMCW}$ is:
\begin{equation}\label{eq:R-velocity}
    R_{C-FMCW} =  \frac{C \cdot Lag}{2F_{s}} - vt
\end{equation}
where $C$ is the speed of sound, $0 \leq Lag \leq 600$ is the number of samples shifted between the transmitted and received signals, and $F_{s}$ is the sampling frequency (the numerical value is specified in our hardware implementation), $v$ is the velocity of the ranged object, and $t$ is the time from the start of the modulation period, $0 \leq t \leq \frac{600}{50000} = 0.012$s.
As reasoned in~\cite{C-FMCW}, when the ranging target is static or moving slowly $R_{C-FMCW}$ and the theoretical resolution upper bound, $\delta R_{C-FMCW}$ are:
\begin{equation}
    R_{C-FMCW} =  \frac{C \cdot Lag}{2F_{s}}
\end{equation}
\begin{equation}
    \delta R_{C-FMCW} = \frac{C}{2F_{s}} =  \frac{343m/s}{2\times 50000} = 3.43 mm
\end{equation}
The improved spatial resolution is critical not only for the original ranging purpose but also for capturing geometries as used in~\theDevice{}.
For the hand pose tracking task, the fingers are sometimes static and sometimes moving.
When considering the ranged object's velocity ($v$), we derive $R_{C-FMCW}$ from eq (\ref{eq:R-velocity}):
\begin{equation}
    R_{C-FMCW} <  \frac{300C}{F_{s}} - \frac{600v}{F_s}
\end{equation}
For reference, in everyday hand movements, the human hand joints' median and max movement speed is less than 10 and 320 deg/s~\cite{handstats}.
Male middle finger (the longest finger) length averages 10.6cm~\cite{fingerlength}, yielding a $v$=0.018m/s and $v$=0.59m/s, which is significantly smaller than the speed of sound.
Therefore, the impact of finger movement on the sensing resolution is very small based on this formula. 
Note, that these numbers are unique to ranging applications and only serve as references for understanding the sensing principle of our system.

\begin{figure}[t]
  \includegraphics[width=\columnwidth]{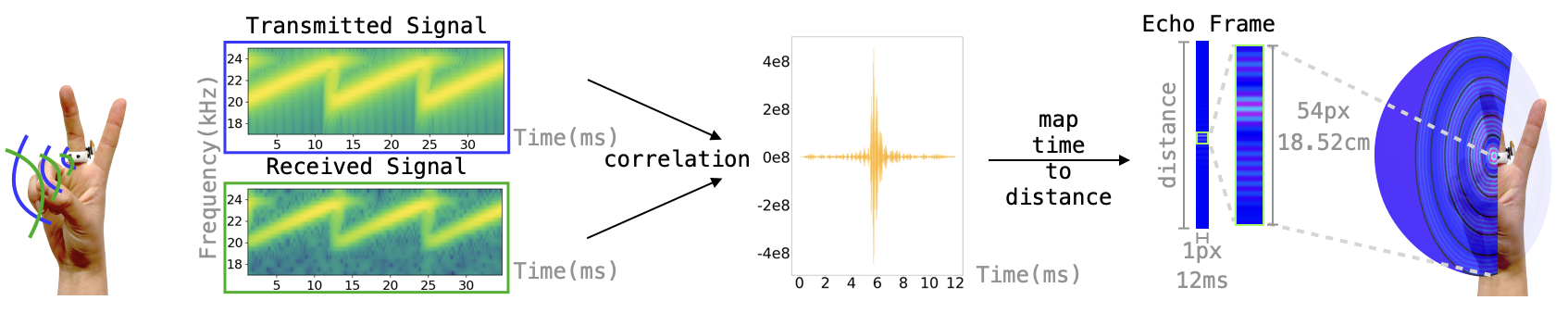}
  \caption{Echo Frame Calculation. The cross-correlation (orange line) between the transmitted  FMCW signal (blue) and the received signal (green) is mapped from the time domain to the distance domain as an echo frame. ~\theDevice{} crops 54 pixels, equivalent to 18.52cm, of the echo frame to analyze hand poses. The black lines in the 3D visualization overlayed on the hand denote a 3cm distance increment. The color of each sphere denotes the summation of reflection strengths from that radius. Note the spectrograms on the left are for visualization purposes, the signals are captured and processed with the time domain.}
  \label{fig:echoframe}
\end{figure}

EarIO~\cite{EarIO}, earables tracking facial movements by measuring subtle skin deformations, first utilized the C-FMCW method to generate echo profiles.
Instead of identifying the length of the strongest reflection strength path which corresponds to the distance between the sensors and the ranged object, Li~\etal{}used the correlation to capture the reflection strengths of different sound travel path lengths.
The calculated echo profiles (detailed below) contain continuous spatial and temporal reflection strengths that denote the complex energy pattern of echos among the different lengths of sound travel paths.
As we reasoned above, these echos are mostly created by the shape of the hand with smaller contribution from the dynamic of the movement.
The echo profiles (bottom 2 rows in Fig.~\ref{fig: echo-gests}) are made up of 1-pixel wide echo frames (Fig.~\ref{fig:echoframe}) stacked along the time axis.
The original echo profile captures the static hand geometries, and the differential echo profile amplifies the hand geometry movements.

\subsubsection{Echo Frame Calculation}\label{sec: ehco frame}
As shown in Fig.~\ref{fig:echoframe}, the ring's speaker emits a Frequency-Modulated Continuous-Wave (FMCW) signal.
The blue box visualizes 3 transmitted frequency sweeps on the left in Fig.~\ref{fig:echoframe}.
The green box on the left in Fig.~\ref{fig:echoframe} visualizes an example signal received by the microphone.
By applying cross-correlation~\cite{C-FMCW} between the transmitted and 5-order Butterworth band-pass filtered (20-24KHz, the same as emitted signals) received signals, we acquire the strengths of the signals at different return/reflection times (orange line in Fig.~\ref{fig:echoframe}).
The strongest correlation (which we centered at 6ms in the correlation graph, like~\cite{EarIO}) at the timestamp $t_{0}$ denotes the direct path between the speaker and the microphone.
We then map the time domain into the distance domain based on the speed of sound.
The hand is small compared to the theoretical sensing range of C-FMCW (2.06m).
The echo frame is 1-pixel wide, equivalent to 12ms (the duration of one frequency sweep), and 600-pixel long (the number of samples per sweep), equivalent to 2.06m. 
The value of each pixel represents the strength of the correlation.
To remove reflections from the environment and focus on sensing the hand, we only use 54 pixels starting from $t_0$, equivalent to 18.52cm, large enough to cover full-hand movements.

\subsubsection{Original \& Differential Echo Profile Calculation}\label{sec: ehco profile}
When we stack echo frames along the time (x-axis), we acquire continuous reflection strengths at different distances. 
The stacked echo frames are called the original echo profile, as shown in Fig.~\ref{fig: echo-gests} middle row.
To capture the changes between echo frames, we calculate the differential echo profile by subtracting the previous echo frame from the current echo frame.
The bottom row in Fig.~\ref{fig: echo-gests} shows the differential echo profile with dynamic range adjustments for visualization purposes.

\subsection{Echo Profiles for ~\theDevice{}}\label{sec:echo-profiles-for-ring}
\begin{figure}[h]
  \includegraphics[width=\columnwidth]{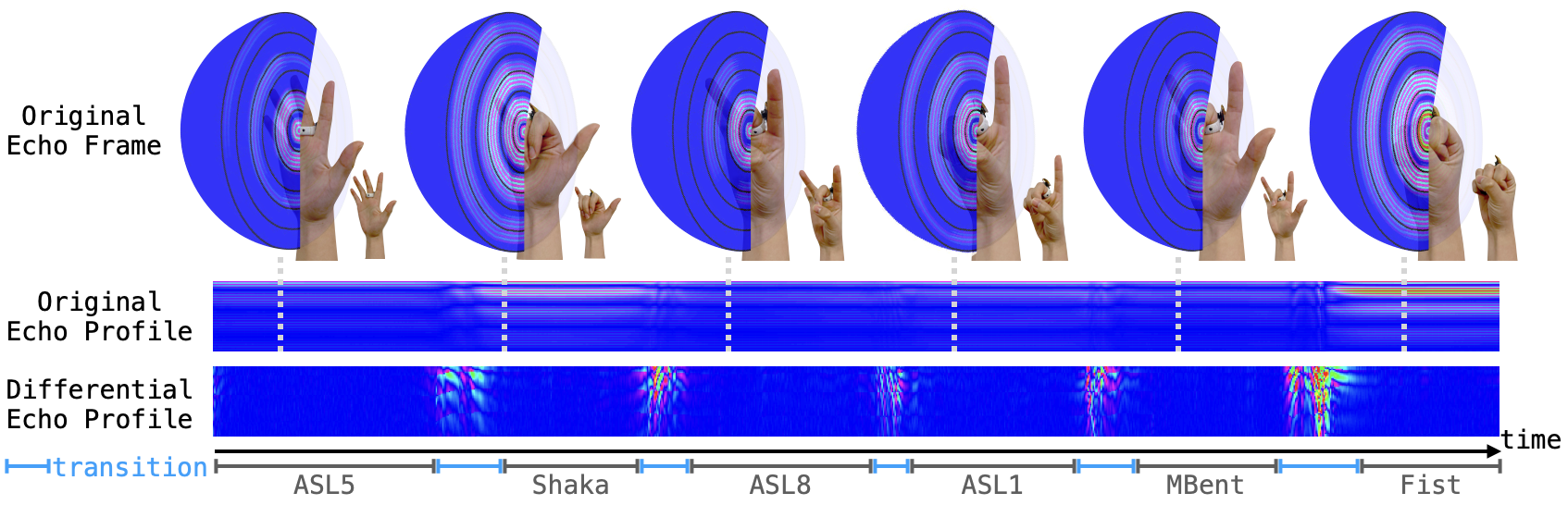}
  \caption{Original and Differential Echo Profile for a sequence of Hand Poses. The black lines in the 3D echo frames visualization overlayed on the hand denote a 3cm distance increment. The color of each sphere denotes the summation of reflection strengths from the travel path lengths of the radius of the sphere. }
  \label{fig: echo-gests}
\end{figure}
Similar sensing approaches have been used for tracking facial movements from earables~\cite{EarIO} and hand poses from a wristband~\cite{EchoWrist}, but additional challenges surface when adapting the technique to hand pose tracking with a ring: (1) the limited physical space allows fewer electronics, reducing the amount of sensed information; and (2) the moving fingers can partially block the sensors, occluding the line-of-sight. We experimented with different sizes of microphones and speakers to minimize the obstruction.

Though there are only two input arrays of information, they complement each other efficiently: the original echo profile encodes the positions of fingers related to poses, and the differential echo profile amplifies the movement of hand geometry (between poses).
Every finger's position and its movements affect the original and differential echo profile, respectively.
Together, they allow us to track 3D hand poses from a single point of instrumentation.
With the omnidirectional speaker, we can abstract the echo frame to be half spheres of various radii centered at the ring (Fig.~\ref{fig: echo-gests}).
The color of the sphere denotes the total strength of reflection from the travel path lengths of the radius of the sphere.
Note, that the echo frame only contains reflection strengths information from different distances, not exact points.
In figure~\ref{fig: echo-gests}, we see that for the ``ASL5'' pose, the flat hand yields little reflection beyond the inner spheres.
``ASL8'' and ``MBent'' are two similar hand poses that only vary in the pose of the thumb, so they share echo frames that are visually similar but distinct to the deep learning model.

Line of sight sensors, that require a clear line between the sensor and the observed object, face the common sensor occlusion issue. 
Though our active acoustic sensing approach shares the obstacle, we found that the ring affords unique placement that turns hand geometrical occlusions (where fingers fully or partially block the ring) into useful and repeatable information. 
``Shaka'' and ``Fist'' are two hand poses where the ring is mostly blocked, so they both have very strong reflections.
However, interestingly, the slight differences in the thumb and pinky finger positions lead to echo frame differences: ``Fist'' has stronger reflections than ``Shaka''.
Later in the evaluation, we investigate the effect of sensor occlusion quantitatively.

Different from many prior data-driven hand pose tracking systems\cite{fingertrak, etherpose, discoband}, one notable strength of our system is that our sensing system is relatively independent of the user (detailed results shown in Sec.~\ref{sec: pose-indepedent}), because our sensing systems rely on the multipath echos reflected by the hands, which are largely determined by the anatomical structure of hands. 
Because human hands are similar anatomically with minor variations in size and shape, they lead to similar echo profiles for the same hand pose across users. 

\subsection{Deep Learning Pipeline}~\label{sec: dl-model}
As described above, echo profiles encode temporal (x-axis) and spatial (y-axis) information of reflection \& diffraction strengths.
With the calculated echo profiles as inputs, our deep learning pipeline predicts the hand pose/gesture with data augmentation techniques.

\subsubsection{Labels for Training/Testing Hand Poses: Ground Truth Acquisition \& Normalization}\label{sec:ground-acquisition}
To capture ground truth for training and evaluation, we use MediaPipe Hands~\cite{mediapipe} that acquire the 3-dimensional (3D) Cartesian coordinates of 21 key points, shown in Fig.~\ref{fig: pose-joints}.
We subtract the wrist key point from the other 20 key points to acquire the relative ground truth joint position with respect to the wrist.
MediaPipe Hands is also used as the ground truth method in ~\cite{etherpose, discoband} to infer 3D coordinates based on RGB images.
We notice the depth prediction of MediaPipe is not precise, but it is consistent under the same lighting conditions.
We include the per-axis performance breakdown in the results section for comparison.
Compared with a depth camera-based ground truth acquisition method, like Leap Motion~\cite{leapmotion}, MediaPipe tracks occluded joints better.
Though the marker-based motion-capturing system is the most accurate, the markers alter the hand geometry and reflection patterns which our acoustic sensing principle relies on. 

To account for hand position and orientation differences across sessions and participants, we normalize the hand orientation.
For each detected hand pose, we find the plane defined by vectors (a) starting from the wrist (joint 0 in Fig.~\ref{fig: pose-joints}) and ending at the index finger metacarpophalangeal (MCP) joint (joint 5 in Fig.~\ref{fig: pose-joints}), and (b) starting from the wrist (joint 0 in Fig.~\ref{fig: pose-joints}) and ending at the little finger MCP joint (joint 17 in Fig.~\ref{fig: pose-joints}).
We then re-align the hand by rotating the palm plane to that of a reference image.
Furthermore, to ensure the same hand size for the same participant, we normalize the hand size based on the physical length measured between the wrist (joint 0 in Fig.~\ref{fig: pose-joints}) and ending at the little finger MCP joint (joint 17 in Fig.~\ref{fig: pose-joints}).

The camera (built-in of Apple Macbook Air 2022) we use with MediaPipe samples at 30 fps, and our ring samples at 83 fps, a much higher frequency.
We synchronize the ground truth with our sensor signals based on timestamps. To minimize real-time inference lagging, we pick the last sensor reading's corresponding hand coordinates as the ground truth for the echo profile window.

\subsubsection{Model Framework}\label{sec:model-framework}
\begin{figure}[t]
\includegraphics[width=\columnwidth]{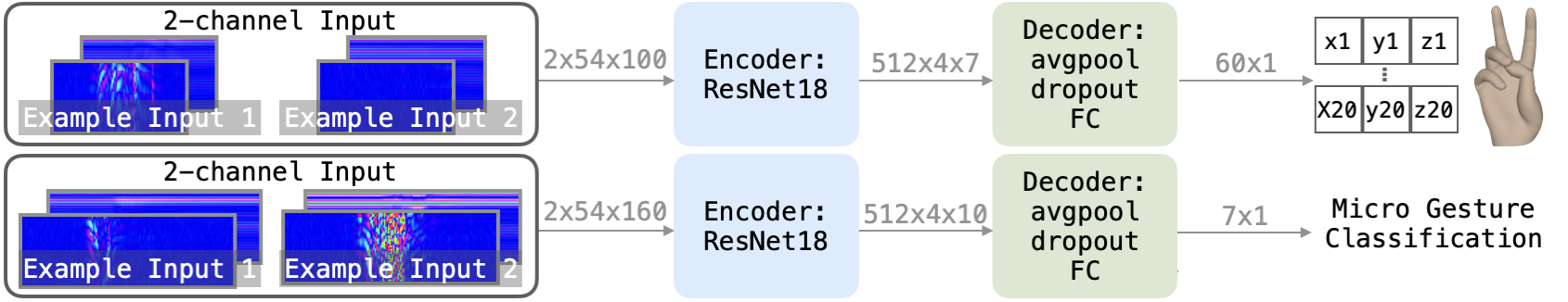}
  \caption{Encoder-decoder Architecture for Hand Pose Tracking and Gesture Classification. Example visualized inputs have the differential echo profile channel in the front and the original echo profile channel in the back.}
  \label{fig:dl-archi}
\end{figure}
The trained deep learning model takes echo profiles as input and outputs either 3D hand joint coordinates or classification labels, depending on the sensing task.
The regression and classification models share similar architectures.
We present detailed comparison results of different algorithms in the Sec.~\ref{discussion: comparative}.

\paragraph{Input: }
The differential echo profile (amplifying hand movements) and original echo profile (capturing static hand poses) are stacked as 2-channel input.
The dimensions of the inputs are $[2 \times 54 \times 100]$ and $[2 \times 54 \times 160]$, equivalent to a 1.2-second (100 pixels) window and a 1.92-second (160 pixels) window for the regression and classification models, respectively.
Note, for the hand pose tracking study data collection, each pose lasts 2 seconds, so some input instances capture the transition between poses (pose tracking example 1 in Fig.~\ref{fig:dl-archi}), while other inputs capture static poses without movements (example 2 in Fig.~\ref{fig:dl-archi}).
Thus, the pose tracking model learns both the static poses and dynamic transitions.
For hand gesture classification data collection, each gesture is performed within a 2-second window, slightly longer than the training instance.
Every collected gesture has its unique input instance.
The input echo profiles are normalized per channel to account for inconsistent magnitudes.

\paragraph{Encoder-Decoder Architecture: }
We adopt ResNet18~\cite{resnet18} as the encoder backbone that outputs a feature vector of size $[512 \times 4 \times 10]$.
The decoder consists of an average pooling layer (pool size=$[1,1]$), a dropout layer ($p=0.8$), and a fully connected layer.
The final fully connected layer either outputs 20 joints x 3 coordinates = 60 coordinates for the hand pose tracking model, or outputs the predicted label for gesture classification models.

\subsubsection{Training Scheme}
Our models are implemented in PyTorch and trained on an NVIDIA GeForce RTX 2080 Ti. For the \textit{hand pose tracking model}, we use the Adam Optimizer with a learning rate scheduler starting at 0.002.
We train the regression model to predict the 20 $\times$ 3 hand joint coordinates using the mean-square error (MSE) loss.
We train the user-independent model with 10 epochs and a batch size of 256.
We fine-tune the model with another 15 epochs and a batch size of 32.
For the \textit{gestures classfication model}, we use the Adam Optimizer with a learning rate scheduler starting at 0.0002.
We train the classification model to predict the gesture label using the cross entropy loss.
The batch size is 32.
We train the user-independent model with 120 epochs and fine-tune the model with another 120 epochs.

\subsubsection{Data Augmentation}
To improve the model's robustness against worn locations, ring orientations, and hand dimension variations, we apply the following in-place data augmentation techniques, further investigated in Sec.~\ref{sec:aug-study}:
\begin{itemize}
    \item Randomness: Applied in both pose tracking and classification models, random $[-5\%,5\%]$ increases at 80\% chance in each pixel of the echo profile that introduces noise to hand geometry.
    \item Vertical Shift: Applied in both pose tracking and classification models, vertical shifts account for the different ring positions relative to the hand. All input echo profiles are randomly vertically shifted by $\pm 3$ pixels, equivalent to $\pm10.02$mm.
    \item Horizontal Shift: Applied only in the classification models, horizontal shifts account for inconsistent reaction times that the participants take to start performing the gesture when requested.
    The input echo profiles are randomly horizontally shifted by $[-13\%,13\%]$ pixels at 80\% chance, equivalent to $\pm0.15$s. Horizontal shifts are not needed for hand pose tracking because the movements are continuous.
\end{itemize}

\section{Evaluation Overview}
To extensively evaluate ~\theDevice{}'s continuous hand pose tracking and recognition performance, we conducted three user studies approved by the Institutional Review Board (IRB): (1) \textbf{the comparison on the performance of different worn fingers}, detailed in Sec.~\ref{sec:study4}, (2) \textbf{continuous hand pose tracking}, detailed in Sec.~\ref{sec:study1}, and (3) \textbf{micro gesture recognition}, detailed in Sec.~\ref{sec:study3}.

The first user study aims to assess the \textbf{impacts of worn finger choice} on tracking performance.
Because the deep learning models are finger-dependent, we selected a finger, the middle finger, for in-depth evaluation.
The sensing tasks were separately evaluated to avoid uncomfortably long studies for the participants and each study had a US\$15 or US\$20 compensation, depending on the study lengths.
The goals of the latter two studies are to assess:
\begin{itemize}
    \item ~\theDevice{}'s \textbf{stability within a participant} so that as the user removes the ring and puts it back on, no additional training data is needed: We address this by asking the participant to remount the ring before each session;
    \item ~\theDevice{}'s \textbf{generalizability across participants} with different hand shapes and hand movement patterns so that no or little training data is needed from a new participant: we address this by evaluating user-independent and fine-tuned models performance; and
    \item ~\theDevice{}'s \textbf{robustness to noise factors} for real-world uses: we address this by testing scenarios with various wrist \& forearm orientations, sounds, movements, and nearby surfaces/objects. 
\end{itemize} 

At the end of each study, the participant completed a survey inquiring about their age, height, weight, and experience with the ring. 
The researcher also measured the lengths and sizes of their fingers. 
In total, we conducted a series of 3 studies with a total of 36 different participants (19 self-identified as male, 17 female, mean age=23.5, std age=3.6), including a variety of hand dimensions, detailed in Table~\ref{table: users}.
Two of the third study participants participated in the first and second studies, respectively.
Note the hand length is the measurement between the tip of the middle finger and the center of the wrist.
\begin{table}[t]
    \small
    \caption{Statistical details of participants' hand dimensions.}
    \begin{tabular}{ | >{\centering\arraybackslash}m{0.8cm} | >{\centering\arraybackslash}m{3cm}| >{\centering\arraybackslash}m{3cm} | >{\centering\arraybackslash}m{3cm}| >{\centering\arraybackslash}m{3cm}| } 
      \hline
           & Middle Finger Ring Size & Hand Length (cm) & Height (cm) & Weight (kg)\\ 
      \hline
      mean & 6.5 & 17.9 & 170.0 & 62.2\\ 
      \hline
      std & 1.6 & 2.0 & 9.12 & 10.4\\ 
      \hline
      max & 10.5 & 22.0 & 189.0 & 93.0\\ 
      \hline
      min& 4.0 & 14.0 & 153.0 & 44.0\\ 
      \hline
    \end{tabular}\label{table: users}
\end{table}

Overall, across all 3 studies, the participants felt comfortable with wearing the ring (Median=4 on the 5-point Likert scale; 1=very uncomfortable, 5=very comfortable).
46\% of the participants could hear the sounds when they performed the gestures, mostly when the hand was near the fist pose, but the sound did not bother them: Median = 4 on the 5-point Likert scale (1=very uncomfortable, 5=very comfortable).

\begin{figure*}[t]
  \includegraphics[width=\textwidth]{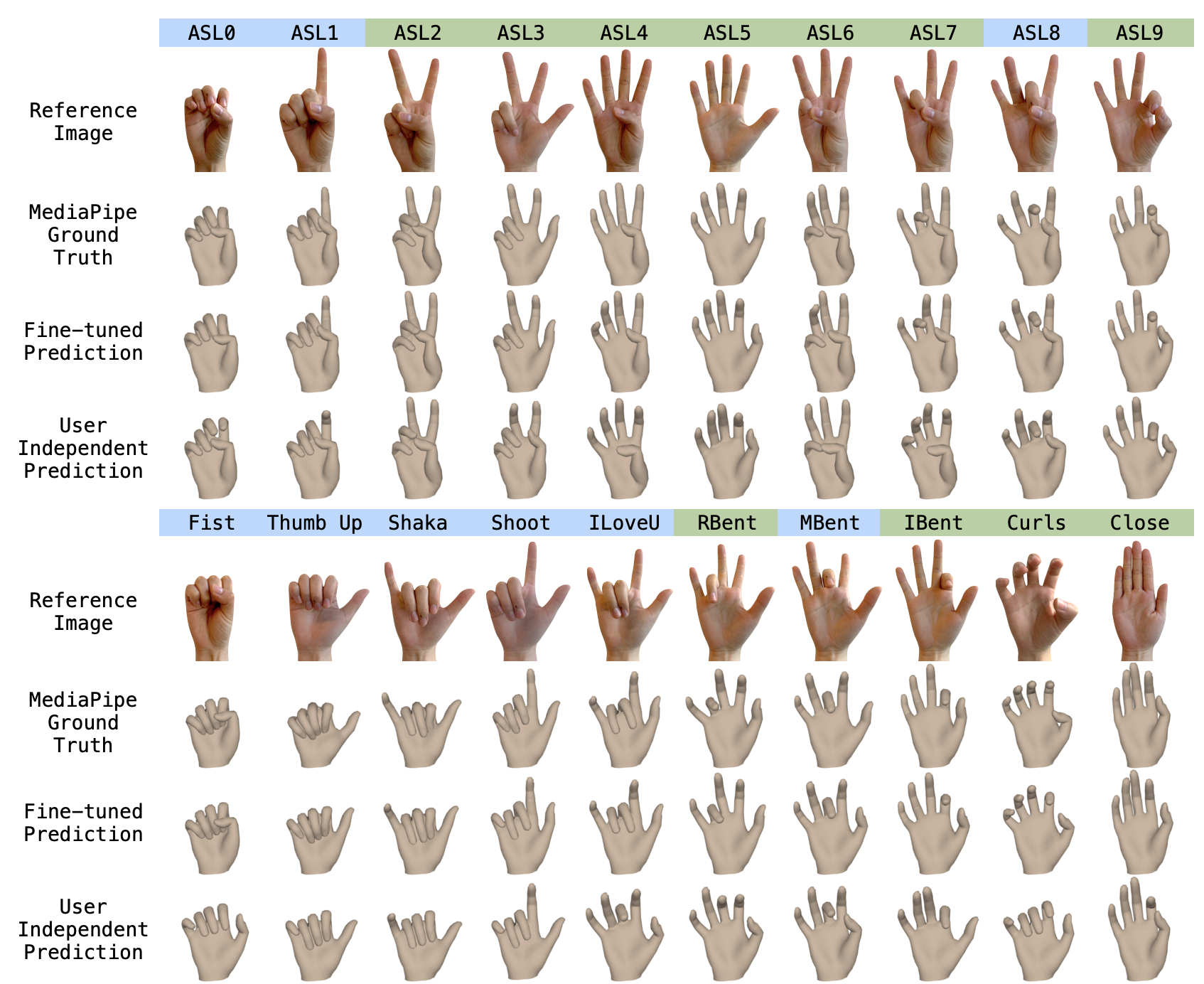}
  \caption{The twenty terminal hand poses evaluated in our hand pose tracking study. Poses are labeled blue or green based on whether the hand geometries occlude the sensor when the ring is worn on the middle finger. The four rows show (1) reference images displayed during the user study, (2) example MediaPipe ground truths of a participant, visualized using MANO~\cite{mano}, (3) example predictions using the fine-tuned model, and (4) example predictions using the user-independent model. (3) and (4) share the same timestamps as (2).}
  \label{fig: poses}
\end{figure*}

\section{User Study 1: Worn Finger Comparison}\label{sec:study4}
As an accessory, a ring can be worn on different fingers based on aesthetic preferences and/or to indicate marital status.
Thus, when designing a sensing ring for wide adoption, it is crucial to investigate the performance of the ring sensing system on various worn fingers. 
It's worth noting that the majority of previous single-ring sensing systems only tested the system on one finger~\cite{RingReview}.
To evaluate ~\theDevice{}'s performance on all possible worn fingers, we conducted a user study with 10 participants.

\subsection{Pose Set}
Because we tested all five fingers on each participant, to ensure a comfortable study duration, we limited to a small pose set and transitions between poses.
We selected (a) 4 isolated individual finger movements: ``ASL4'' (equivalent to bending the thumb), ``RBent'', ``MBent'', and ``IBent`` in Fig.~\ref{fig: poses}; and (b) 5 compound finger movements: odd ASL digits between 0 and 9.
Between each 2 poses, the hand returns to a neutral position (``ASL5'').

\subsection{Procedure}
\subsubsection{Study Setup.}
We conducted a user study with 10 participants.
The user study was conducted in an experimental room on a university campus. At the beginning of the study, one researcher explained the study task and the user interface displayed on a laptop monitor. Participants sat in desk chairs during the study with their elbows on the table to ensure that the camera could capture their hands for ground-truth purposes.

A smartphone (Xiaomi Redmi Note 10 Pro) was used to receive and save BLE-transmitted acoustic data from the ring. Participants were instructed to independently wear the ring on their right middle finger before each session, following the experimenter's guidance for alignment. Ground truth data for hand pose was collected using the built-in laptop camera (Apple Macbook Air, 2022) positioned about 55 cm away from the hand on the table and powered by MediaPipe.
The experimenter adjusted the camera angle to ensure that the palms stayed parallel to the camera for reliable ground truths. 
The study lasted about 90 minutes.

\subsubsection{Data Collection Sessions.}
Each participant underwent 21 sessions: 1 practice session followed by 4 sessions for each finger in a randomized order (1 practice + 4 sessions * 5 fingers = 21 sessions).
Participants were instructed to independently wear the ring on the selected finger before each session, following the experimenter's guidance for alignment: centering the inner edge of the speaker with respect to the selected finger.
Within each session (2.7min), each pose is performed 9 times in a randomized order: 9 $\times$ 9 = 81 terminal poses.
A reference pose image was displayed on the monitor for 2 seconds for each pose as the visual stimuli.
Within 2 seconds, the participant's hand leaves the neutral (``ASL5'') position to the referenced pose and returns.
For each finger, we collect 4 sessions $\times$ 2.7 min = $10.8$ min of data containing 4 sessions $\times$ 81 terminal poses = $324$ terminal pose instances.

\subsection{Results}
\paragraph{Evaluation Metrics}
We adopt the mean per-joint position error MPJPE to be our quantitative evaluation metric: the mean Euclidean distance errors of all 20 relative (to the wrist) joint positions.
Note, MPJPE measures distance errors so it depends on the hand size.
An alternative evaluation metric is the mean angular error, but due to MediaPipe Hands' unreliable depth predictions, we chose MPJPE.
To account for MPJPE's dependency on the hand size in the user-independent models, we normalize the predicted hand with the participant's physical hand size, as described in Sec.~\ref{sec:ground-acquisition}.
In real-world uses, it is an additional step a new user needs to do when receiving the device.
For fine-tuned user-dependent models, such information is no longer needed as the model quickly learns the hand dimension as shown in the convergence of MPJPE with only 2.67 min of data in Fig.~\ref{fig: pose-UIUD}(c).

We compare~\theDevice{}'s performance on each finger with user-independent models and the fine-tuned user-dependent models.
For each finger, to evaluate the user-independent performance, we use the leave-one-participant-out (LOPO) cross-validation (97.2min of training data from other participants' same finger); and to evaluate the fine-tuned user-dependent performance, we fine-tune the user-independent model with data from the first three sessions (8.1min of fine-tuning data) and test on the last session.
In addition to the MPJPE of the entire hand, we further break down the results into the MPJPEs of each finger to analyze whether the worn finger affects the tracking accuracies of individual fingers.
\begin{figure*}[t]
  \includegraphics[width=\textwidth]{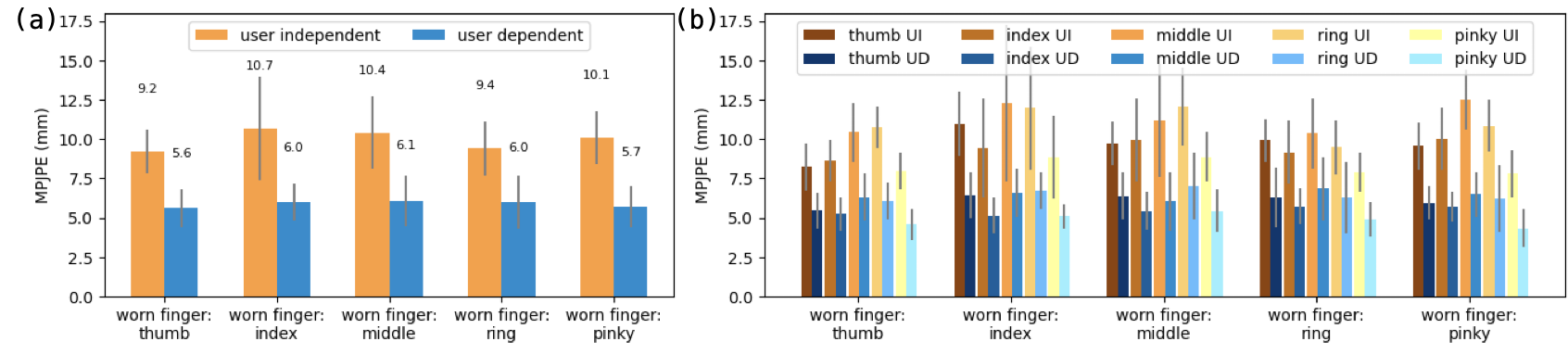}
  \caption{Worn Finger Comparison Study Results. (a)~\theDevice{}'s tracking performance when worn on different fingers. (b) Tracking performance breakdown on each tracked finger when worn on different fingers. The legend denotes the tracked finger and model type. For example, ``thumb UI'' and ``thumb UD'' are the user-independent and the user-dependent tracking errors on the thumb, respectively. Error bars in this figure represent the standard deviation.}
  \label{fig: figners-results}
\end{figure*}
~\theDevice{} has similar performances when worn on all five fingers.
The results are detailed in Fig.~\ref{fig: figners-results}(a).
For user-independent models, the errors range from 9.2mm (thumb) to 10.7mm (index finger).
For user-dependent models, the errors range from 5.6mm (thumb) to 6.1 mm (middle finger).
We break down the individual tracked finger's performance (Fig.~\ref{fig: figners-results}(b)): the worn finger generally tracks itself better than other fingers do.
The multivariate analysis of variance (MANOVA) with model type and worn fingers as independent variables and individual tracked finger errors as dependent variables showed that the worn finger has a significant effect (F$_{5,93}$=3.35, p=0.008<0.05) on individual tracked finger errors.
We then follow up the MANOVA test with individual ANOVA tests for each tracked finger, we find the thumb (F$_{4,36}$=5.76, p=0.001<0.05), ring finger (F$_{4,36}$=3.35, p=0.020<0.05), and pinky finger  (F$_{4,36}$=3.58, p=0.015<0.05) depend on the worn finger, and the other tracked fingers do not.
But in most practical cases, the hand is tracked as a whole.
A two-way repeated ANOVA test on model types and worn fingers showed that the worn finger factor is not statistically significant (F$_{4,36}$=2.26, p=0.08).

In summary, while~\theDevice{} has comparable performance across all worn fingers for full-hand tracking, it tracks individual fingers differently based on the worn finger.
The result implies that for a general-purpose hand-pose tracking solution, the worn finger location does not matter: unlike most prior works~\cite{RingReview}, users are free to choose which finger to wear the ring based on their own preferences without performance limitation.
However, if the application demands higher tracking accuracies on certain fingers (e.g., thumb and index finger for pinch detections), the worn finger location can alter the performance.
We acknowledge that our sample size of 12 (participants) is small for statistical analysis, but in practice, designers and researchers can pick the worn location based on their needs.

\section{User Study 2: Continuous Hand Pose Tracking}\label{sec:study1}
In this section, we detail the procedures and findings from the continuous hand pose tracking study.

\subsection{Pose Set}
To capture a wide range of hand poses and movements, we select a set of 20 terminal poses, shown in Fig.~\ref{fig: poses}.
Informed by prior works~\cite{opisthenar, etherpose, discoband, fingertrak}, our pose set includes all 10 American Sign Language (ASL) digits evaluated in Opisthenar\cite{opisthenar} and all 11 poses evaluated in prior work ~\cite{etherpose}.
\newline
\subsubsection{Natural Transition between Poses}~\label{sec:pose-transitions}
Just like our first user study, many prior works related to continuous hand pose tracking~\cite{fingertrak, WR-Hand, Back-hand-pose}, necessitate that participants revert to a neutral hand pose between transitioning to different poses. However, this requirement deviates from many real-world scenarios where individuals seamlessly move from one pose to another without returning to a neutral position.
Tracking poses without the need to return to a neutral pose poses significantly greater challenges for data-driven pose-sensing systems, because it exponentially increases the number of potential hand shapes and poses between two target poses, demanding a much larger volume of training data. We deliberately opted for the more challenging task of instructing users to perform hand poses without reverting to the neutral pose.

\subsubsection{The Impact of Sensor Occlusion by Fingers}

To investigate the impact of sensor occlusion by the fingers, we divide the pose set based on whether the ring, worn on the middle finger, is partially (\textit{e.g.} ``ASL8'') or even fully (\textit{e.g.} ``Fist'') occluded by the hand geometry: (a) NO-poses, no-occlusion poses colored in green, and (b) WO-poses, with-occlusion poses colored in blue in Fig.~\ref{fig: poses}.
Further, we purposely include terminal poses with similar occlusions (\textit{e.g.} ``Shaka'' v.s. ``Thumb Up'' and ``Shoot'' v.s. ``ILoveU'') to probe~\theDevice{}'s performance with occlusions.

\subsection{Study Procedure}
\subsubsection{Study Setup}
We first recruited 12 participants on the university campus for the main study and then recruited another 6 participants for the follow-up study on robustness to nearby surfaces and objects (Sec.~\ref{sec:surface-results}). 
The study setup is similar to that in the first study, except all data are collected with the ring worn on the middle finger.

\subsubsection{Data Collection Sessions}
Each participant from the main study underwent 25 sessions, and within each session, they were tasked with performing three sets of gestures: 1) all 20 poses; 2) 11 No-poses; and 3) 9 WO-poses. The order of these terminal poses, within each set of poses, was randomized. Between two terminal poses, transition poses are also recorded for evaluation. To assist participants in performing the target poses, a reference pose image was displayed on the monitor for 2 seconds for each pose, accompanied by a progress bar as a visual stimulus. We did not strictly enforce hand pose transition speeds. These 25 sessions can be categorized into three sections:

1) Practice Section (Sessions 1-2, 2.56 min each): The first two sessions were designed as practice sessions to help the participants get familiar with the target poses and the data collection interface. The data from practice sessions were not used in the evaluation.
    
2) Major Testing Section (Sessions 3-14, 2.56 min each): During the subsequent 12 sessions, the procedure closely mirrored that of the practice sessions. In each of these sessions, each participant performed 2$\times$(20+11+9) = 80 poses. In total, one participant performed 12$\times$2$\times$(20+11+9) = 960 terminal poses in the major testing section. 
    
3) Robustness Testing Section (Sessions 15-25, 1.28 min each): In the robustness testing section, we sought to evaluate our system under 11 distinct conditions that could potentially affect its performance in real-world settings. Each session was specifically designed to assess one condition.  Unlike the previous sessions, the participant only performs the three sets of gestures once per session, leading to 1$\times$(20+11+9) = 40 terminal pose instances per session. These 11 sessions represent the following scenarios: 

\begin{enumerate}
    \item Pose-Neutral-Pose: Between each prompted pose, the hand returns to a neutral position (``ASL5'').
    \item Uncontrolled Hand Movement: Participants moved their hands freely in front of the screen while performing the poses, as long as their hands stayed parallel and inside the camera view for ground truth acquisition.
    \item Environmental Acoustic Noise (Music Playing): Participants played music of their choice using their phone's speaker at a volume they typically listen to music at.
    \item Talking: Participants chose to talk either with the experimenter or monologue throughout the session.
    \item Radial Deviation: Wrist stayed in the radial deviation orientation.
    \item Ulnar Deviation: Wrist stayed in the ulnar deviation orientation.
    \item Flexion: Wrist stayed in the flexion orientation.
    \item Extension: Wrist stayed in the extension orientation.
    \item Neutral: The forearm stayed in the neutral orientation.
    \item Supination: The forearm stayed in the supination orientation.
    \item Hand-Down: The right hand pointed downwards next to the participants' legs.
\end{enumerate}

For the main study, due to a hardware malfunction, 1 participant's data was damaged and this participant was invited to participate in the user study again.  In total, we collected 531 mins of pose (each participant 44.25 min x 12 participants) containing 22080 (12 participants x (960+440) = 22080) terminal pose instances.
For the follow-up study, there are 10 major testing sessions with the same setup.
For the robustness testing sections, the scenarios (Fig.~\ref{fig: pose-surface}) are changed to the following to study the effect of nearby objects/surfaces while ensuring reliable ground truth acquisition:
\begin{enumerate}
    \item Acrylic sheet at 36 cm away from the hand.
    \item Acrylic sheet at 36 cm away from the hand with uncontrolled hand movements: Similar to the second scenario in the main study above, the participants moved their hands freely in front of the acrylic sheet. As a result, the distance between the hand and the acrylic sheet varies throughout the session.
    \item Acrylic sheet at 24 cm away from the hand.
    \item Acrylic sheet at 12 cm away from the hand.
    \item Plastic box to the side of the hand.
    \item Box covered with cotton fabric to the side of the hand.
    \item Box covered with copper tape to the side of the hand.
    \item Plushie to the side of the hand.
\end{enumerate}

For the follow-up study, due to a hardware malfunction, 1 participant's data was damaged and we recruited an additional participant.

\subsection{Results}
The evaluation metric is the same as the first study: mean per-joint position error (MPJPE).
For the noise factors, we first fit a linear mixed-effects model with the model type and the robustness testing condition as independent variables, and then we follow with a post-hoc Dunnett's test for comparisons with a control (i.e., the major testing condition).
Note we do not evaluate within-session performance because~\theDevice{} generalizes well with remounting.
\begin{figure*}[t]
  \includegraphics[width=\textwidth]{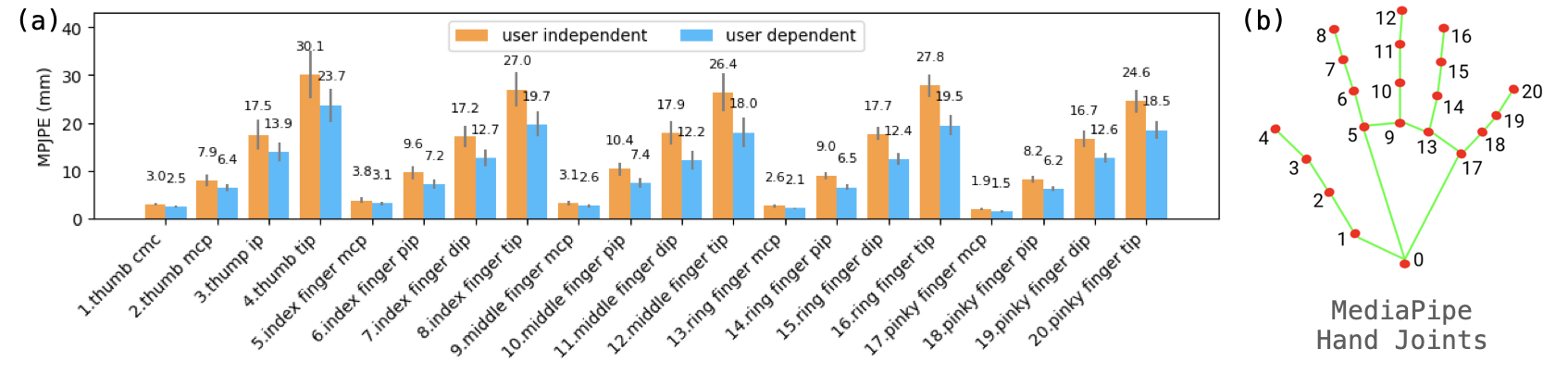}
  \caption{Joint errors Breakdown. The label indices in (a) map to the mediapipe joint labels in (b) that show MediaPipe hand joint labels. Error bars in this figure represent the standard deviation.}
  \label{fig: pose-joints}
\end{figure*}
\subsubsection{User-Independent Performance}\label{sec: pose-indepedent}

We use a leave-one-participant-out (LOPO) cross-validation to simulate when the new user does not provide any calibration data.
For each participant, in addition to the data from the other 11 participants (43.97 min x 11 = 8.06h), we also add data collected from 5 researchers (9.96h).
In total, 18.02 hours of training data from 16 hands are used.

Overall, we achieved a mean MPJPE of 14.1 mm (SD=5 mm) across 12 participants.
The impact of sensor occlusion is shown in Fig.~\ref{fig: pose-UIUD}(a). 
The error distribution of joint distance errors is shown as the orange line in Fig.~\ref{fig: pose-UIUD}(b).
The breakdown of the errors in different axes are: x (left/right of the palm) = 5.8 mm, y (up/down of the palm) = 8.0 mm, and z (front/back of the palm) = 6.7 mm.
The individual joint error breakdown, orange bars in Fig.~\ref{fig: pose-joints}(a), show that the joints that are further away from the wrist exhibit larger errors, not surprisingly.

\subsubsection{Fine-Tuned User-Dependent Performance}
\begin{figure*}[t]
  \includegraphics[width=\textwidth]{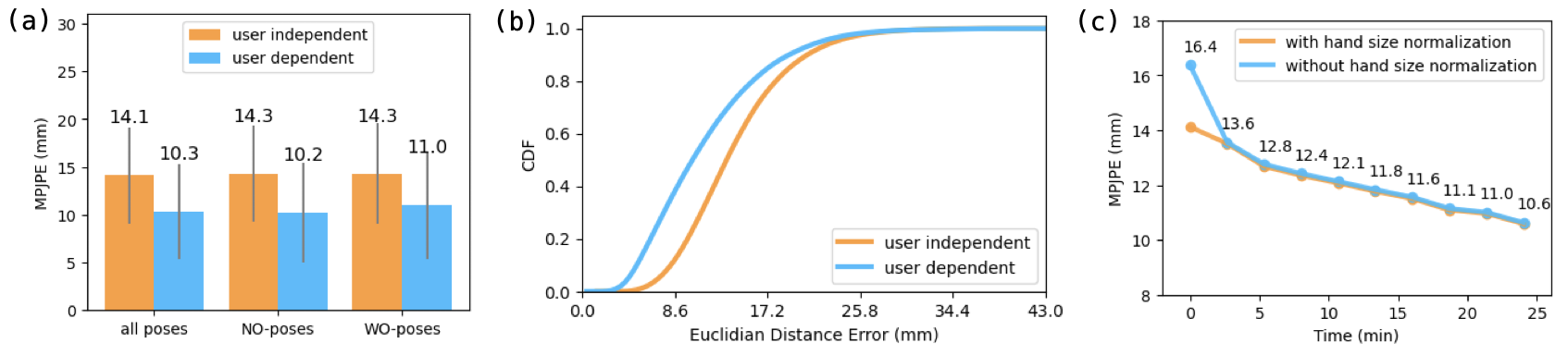}
  \caption{User Inpendent and User-Dependent Fine-Tuned Hand Pose Tracking Results. (a):~\theDevice{}'s performs well on both poses that fully/partially occlude the sensor (WO-poses) and those that do not (NO-poses). (b): Overall Per-Joint Error distribution. (c): With more training data from a new user, the sensing performance increases. The model quickly learns the physical dimension of the hand in just 2.5 min. Error bars in this figure represent the standard deviation. }
  \label{fig: pose-UIUD}
\end{figure*}

Human hands vary in size, shape, and movement patterns across different people.
Based on the user-independent models trained above, we fine-tune the models with different amounts of training data from the new participant.
Additional training data from a new participant improves the tracking performance, as shown in Fig.~\ref{fig: pose-UIUD}(c).
With 26.7 min of additional fine-tuning data, the overall MPJPE decreases to 10.3 mm, 3.8 mm smaller than the user-independent model.
Example hand predictions are included in the 3rd rows of Fig.~\ref{fig: poses}.
In Fig.~\ref{fig: pose-UIUD}(b), the blue line shows the error distribution of joint distance errors.
The breakdown of the per-axis errors decrease to x = 4.6 mm, y = 5.8 mm, and z = 4.8 mm, which are 1.23 mm, 2.18 mm, and 1.9 mm smaller than the user-independent model: the improvements are consistent with movement ranges along each axis.
The blue bars in Fig.~\ref{fig: pose-joints}(b) show the per-joint improvement: the performance increases more for the joints that are further away from the wrist.
As illustrated in Fig.~\ref{fig: pose-UIUD}(c), the quick MPJPE convergence between with and without normalization shows that the model quickly learns the physical hand dimension with just 2.67 min of data.
Further, the slopes in Fig.~\ref{fig: pose-UIUD}(c) decreases as more data is added, but they do not yet flat out.
This indicates that with more fine-tuning data, there is still room for improvement.
\subsubsection{The Impact of Sensor Occlusion by Fingers}
As mentioned above, to investigate the impact of sensor occlusion by the fingers, we divide the pose sets NO-poses (No occlusion) and WO-poses (fully/partially occlude the sensor) based on whether the ring is partially or fully occluded by the hand geometry.
Fig.~\ref{fig: pose-UIUD}(a) shows the tracking performance on the two pose sets.
We performed a two-way repeated ANOVA test, with the model type and occlusion conditions as the independent variables, and did not find statistically significant effects (F$_{1,11}$=1.03, p=0.33) of the pose type on the tracking performance, and there were no interactions (F$_{1,11}$=1.36, p=0.27) between the user- independent/dependent models and NO-/WO- poses.

\subsubsection{Robustness to Wrist \& Forearm Orientations}
A clear benefit of using the ring to track hand poses, compared with wrist-worn devices, is the minimized discrepancy between the hand orientation and the sensor orientation.
The user-independent model is trained with both neutral wrist position data and all other wrist orientation data from other users. The user-dependent model is only fine-tuned with data collected in the neutral wrist orientation.
We then test the models with data collected from sessions with the 6 extreme orientations as shown in Fig~\ref{fig: pose-robust}(b).
The wrist orientation columns in Fig.~\ref{fig: pose-robust}(a) show that, except for supination (t$_{253}$=4.30, p=0.003$<$0.05 from the Dunnett's test), the user-independent model performs similarly across all orientations: 0.55mm decrease in performance without statistical significance (p-value range: 0.36-0.99).
The user-dependent models have small decreases (mean=2.4 mm) in performance and all but the ulnar wrist orientation (t$_{253}$=2.52, p=0.10) have statistical significance (p-value range: 0.000-0.001).
For the supination orientation, MPJPE has big increases for both the user-independent model and the fine-tuned models.
During the user studies, this orientation was hard to perform to keep the palm parallel to the screen. 
In the detected ground truths, we also noticed many frames with visually incorrect hand poses.
Thus, the increase in errors may not accurately depict our system's performance.

Furthermore, we tested the performance when the hand is down on the side of the body (``Hand Down'' column in Fig.~\ref{fig: pose-robust}(a)), a more natural interaction site.
The user-independent model performed similarly (t$_{253}$=-0.70, p=0.97) to that in the training orientation, but the fine-tuned model (t$_{253}$=5.73, p$<$0.001) performance degraded to that of the user-independent model.
In summary, we conclude that the wrist\&forearm orientations have little impact on the more generic user-independent model, but have a performance degradation for the fine-tuned model.

\subsubsection{Robustness to Sounds}
\begin{figure*}[t]
  \includegraphics[width=\textwidth]{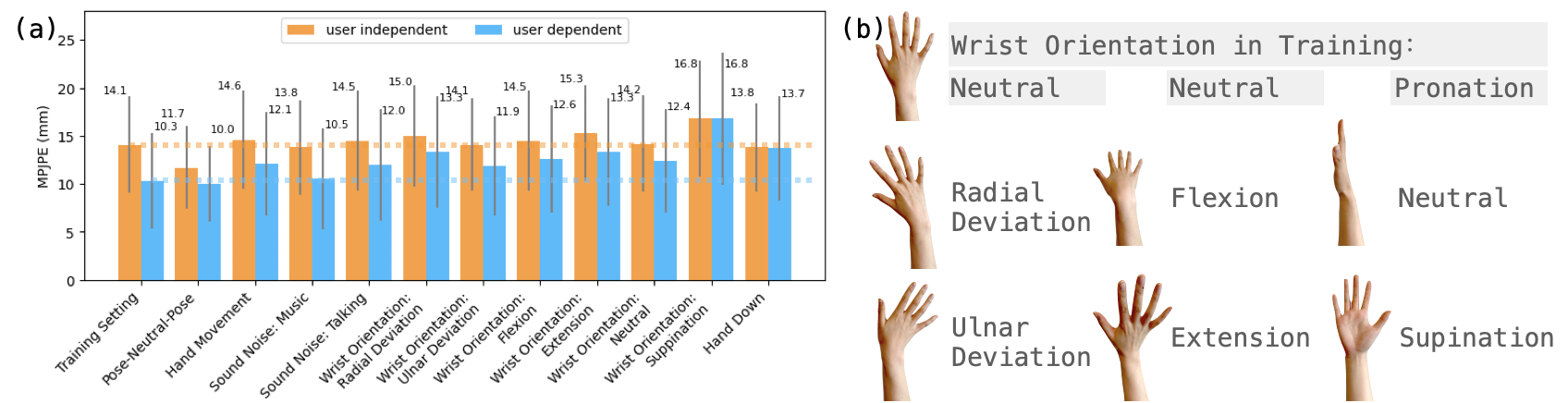}
  \caption{~\theDevice{}'s Robustness to Noise Factors. (a) Using models trained with data collected in the same setting, the testing results on data collected in various settings show that~\theDevice{} is robust to many noise factors. (b) Wrist Orientations. Error bars in this figure represent the standard deviation.}
  \label{fig: pose-robust}
\end{figure*}

Because~\theDevice{} leverages active acoustic sensing, it is necessary to evaluate the performance with environmental sounds.
The two selected sound noise scenarios are visualized in the ``Music'' and ``Talking'' columns in Fig~\ref{fig: pose-robust}(a).
The mean amplitude levels (in dBFS) in the audible range (20-16000Hz) calculated from the microphone recordings without sound noise, with ``Music'', and with ``Talking'' are -54.6, -53.6, and -51.0.
For reference, we used a sound level meter app\footnote{https://www.cdc.gov/niosh/topics/noise/app.html} on an iPhone which colocates with the ring.
In the experiment room when with only ambient noise, we measured 38 dB(A) with the app and -55.06 dBFS with the ring; when the computer plays music at 68cm, 45cm, and 23cm away from the ring\&phone, we measured 47 dB(A)/-55.02 dBFS, 53 dB(A)/-55.02 dBFS, 58 dB(A)/-52.94 dBFS.
The measured dBFS change is relatively small as the noise sound source is much further away from the microphone on the ring than the speaker on the ring.

With the participant's choice of music played at their preferred volume, we do not see performance degradation for both the user-independent (t$_{253}$=-0.81, p=0.94) and -dependent (t$_{253}$=0.114, p=1) models when compared to the result in the training setting shown in Fig~\ref{fig: pose-robust}(a).
However, when the participant talks as they perform the gestures, we see a small degradation without statistical significance: 0.4mm and 1.7mm increase in MPJPEs for the user-dependent (t$_{253}$=0.39, p=1) and user-independent (t$_{253}$=2.6, p=0.08) models, respectively.
We are unsure about the cause of the different effects brought by music and talking, but regardless, the system still performs well with environmental sounds.
\subsubsection{Robustness to Movement}\label{sec:robust-movements}
Since people's hands move when they speak, gesture, and interact with spatial interfaces, we evaluate our system under these scenarios. 
The ``Hand Movement'' column in Fig.~\ref{fig: pose-robust}(a) shows that the movements slightly harm the performance by 0.5mm for a user-independent model (t$_{253}$=0.55, p=0.99) but yields an additional 1.8mm increase for the user-dependent model (t$_{253}$=2.91, p=0.03$<$0.05).

\subsubsection{Robustness to Nearby Objects}\label{sec:surface-results}
\begin{figure*}[t]
  \includegraphics[width=\textwidth]{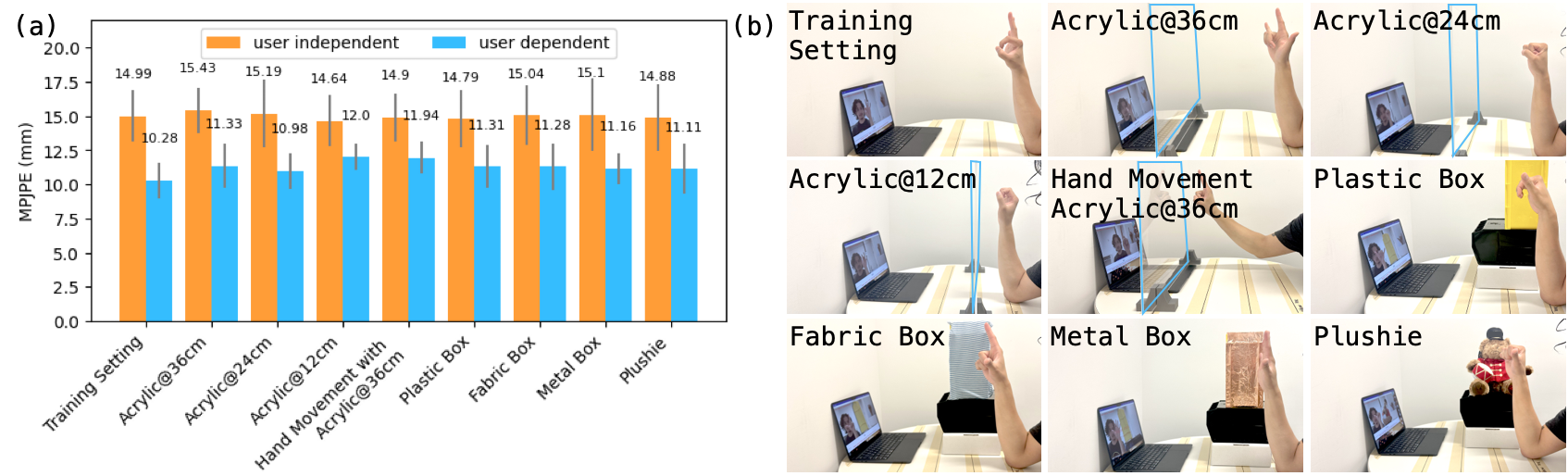}
  \caption{~\theDevice{}'s Robustness to Nearby Surfaces and Objects. (a) Using models trained with data collected in the same setting, the testing results on data collected in various settings show ~\theDevice{}'s performance with nearby surfaces and objects. (b) Training and Tested Settings. The acrylic sheets are highlighted in light blue. Error bars in this figure represent the standard deviation. }
  \label{fig: pose-surface}
\end{figure*}

In addition to finger occlusions, the sensor could be occluded by nearby surfaces and objects(\textit{e.g.}, pants, and desks).
Fig.~\ref{fig: pose-surface}(b) shows our tested scenarios with objects placed near the hand, without interfering with our vision-based ground truth acquisition method.
Note that the laptop for data collection is placed about 55 cm away from the hand, and we crop the echo profiles at 18.52 cm.
With data collected from the six participants in the follow-up study, we evaluate the models' robustness and show the results in Fig.~\ref{fig: pose-surface}(a).

For the user-independent model (trained with 5 other participants' data), we do not find statistical significance (from Dunnett's test) between the training setting and the testing setting: the smallest p-value is 0.94.
For the user-dependent model (fine-tuned with 9 sessions of 23min data), the ``acrylic@12cm'' scenario is significantly different from the training setting (t$_{85}$=2.80, p=0.04$<$0.05) with a 1.72 mm increase in error.
The next largest performance degradation is ``hand movement with acrylic@36cm'' scenario with 1.66 mm increase in error with a trend towards significance (t$_{85}$=2.70, p=0.05).
This finding is consistent with the results from ``hand movement'' without the acrylic sheet in the front, from Sec.~\ref{sec:robust-movements}.
For other scenarios, p-values range from 0.43 to 0.76.
Objects directly in front of the hand and within the echo profile sensing range harm the user-dependent performance.
For other nearby objects, they have little effect on the tracking performance.
Since our sensing range is small, the limitation caused by object occlusion is less prevalent than that from vision-based solutions.

\begin{table}[t]
    \centering
    \caption{Continuous hand pose tracking performance comparison. In the ``Pose/Gesture Set'' column, ``gestures'' refer to when a neutral hand pose is required between transitioning to different poses, and ``poses'' refer to such neutral pose is not required between transitions. In the error columns, ``SD'' stands for session dependence, ``UD'' stands for user dependence, and ``UI'' stands for user independence. All errors refer to mean per-joint position error (MPJPE).}
    \begin{tabular}{|c|c|c|c|c|c|}
        \hline
        System & Devices & Pose/Gesture Set & \begin{tabular}[c]{@{}c@{}}SD \\ Error \\ (mm)\end{tabular} & \begin{tabular}[c]{@{}c@{}}UD \\ Error \\ (mm)\end{tabular} & \begin{tabular}[c]{@{}c@{}}UI \\ Error \\ (mm)\end{tabular}\\
        \hline
        ssLOTR, 2022~\cite{learning-on-the-rings} & 5 rings and a wristband & free movements & \textemdash & 6.55& \textemdash\\
        \hline
        WR-Hand, 2021~\cite{WR-Hand} & an armband & 11 gestures + 3 free motions & \textemdash & \textemdash & 18.57\\
        \hline
        FingerTrak, 2020~\cite{fingertrak} & a wristband & 19 gestures & 12 & 27.2 & \textemdash \\
        \hline
        EchoWrist, 2024~\cite{EchoWrist} & a wristband & 18 gestures & \textemdash & 4.81 & 12.2 \\
        \hline
        DiscoBand, 2022~\cite{etherpose} & a wristband & 10 poses & 11.69 & 17.87 & 19.98\\
        \hline
        Z-pose, 2023~\cite{z-pose} & a ring & 10 poses & 8.5& \textemdash & \textemdash \\
        \hline
        \textbf{\theDevice{}} & a ring & 20 poses & \textemdash & 10.3& 14.1\\
        \hline
    \end{tabular}
    \label{table:tracking}
\end{table}

\subsection{Takeaways}
To situate~\theDevice{}'s continuous hand pose tracking performance with other wearable solutions, we compare its performance with that of prior works in Table~\ref{table:tracking}. Note that this comparison may not be completely fair, as each system used different hardware and different datasets. Because, unlike the vision-based hand-pose tracking systems which use the same benchmark datasets, creating such benchmark datasets is nearly impossible for wearables with customized hardware and varying sensing principles. Nonetheless, the aim of this comparison is to present results and assist readers in better understanding the positioning of our system relative to prior works. 

Because~\theDevice{} is the first single-ring system that tracks continuous hand pose across sessions and users, we compare the performance with multi-ring systems~\cite{learning-on-the-rings}, armbands~\cite{WR-Hand} and wristbands~\cite{fingertrak, discoband}, in addition to single-ring systems~\cite{z-pose}.
Although we can directly compare the MPJPEs, they are heavily affected by the pose/gesture sets (detailed in Sec.~\ref{sec:pose-transitions}), the evaluation condition, and the amount of training data. Regardless,~\theDevice{} demonstrates promising session-independent and user-independent performances when compared with other single-device systems.
~\theDevice{} falls short of the multi-device system that contains 5 rings and a wristband but greatly alleviates the burden of wearing a sensing ring on every finger.

Our user-independent model generalizes well across different hand sizes/shapes and is robust to noise factors.
The fine-tuned model indeed benefits from user-specific data. 
Both models show resilience to the noise factors, but compared with the user-independent model, the fine-tuned model is more vulnerable to noise factors.
One reason for this is that the user-independent model is trained with some "noise" (using both 12 long sessions and 11 short sessions), but the fine-tuned model is not fine-tuned with "noise" (only using the 12 long sessions), so including more diverse training data for training fine-tuned models will be helpful. 
In this evaluation, we isolate the "noise" factors for controlled evaluation.
Future works on in-the-wild studies will help us understand the system better, but it is challenging to acquire ground truth data in the wild.

\section{User Study 3: Thumb-to-Index Micro-Gesture Recognition}\label{sec:study3}
The previous showed ~\theDevice{} continuously track the hand poses effectively.
In this study, we evaluate ~\theDevice{}'s performance on tracking fine-grained micro-gestures which are more discreet, intuitive, and natural~\cite{gheran2018gestures}.

\begin{figure*}[t]
  \includegraphics[width=\textwidth]{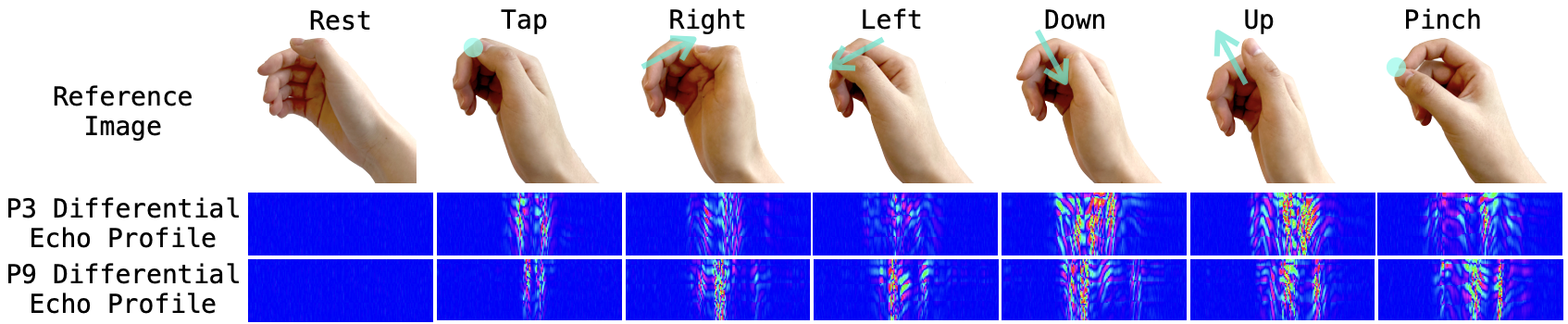}
  \caption{The seven micro thumb-to-index gestures evaluated in our user study. Top row: reference images similar to those displayed during the user study. Middle row: Example differential echo profiles from participant 3. Bottom row: Example differential echo profiles from participant 9.}
  \label{fig: microgests}
\end{figure*}

\subsection{Thumb-to-Index Micro-Gesture Set}
Among various micro gesture sets~\cite{etherpose, thumbtrak, handsense}, we chose thumb-to-index micro-gestures~\cite{pyro, Thumb-In-Motion, efring} due to their ease of performance and social acceptance~\cite{singlehandmicro}.
Shown in Fig.~\ref{fig: microgests}, we chose 7 gestures, including a ``Rest'' no-gesture class.
Similar to that in the previous study, each gesture started from the ``Rest'', moved the thumb and index finger, and returned to the ``Rest'' position.
\begin{figure*}[t]
  \includegraphics[width=\textwidth]{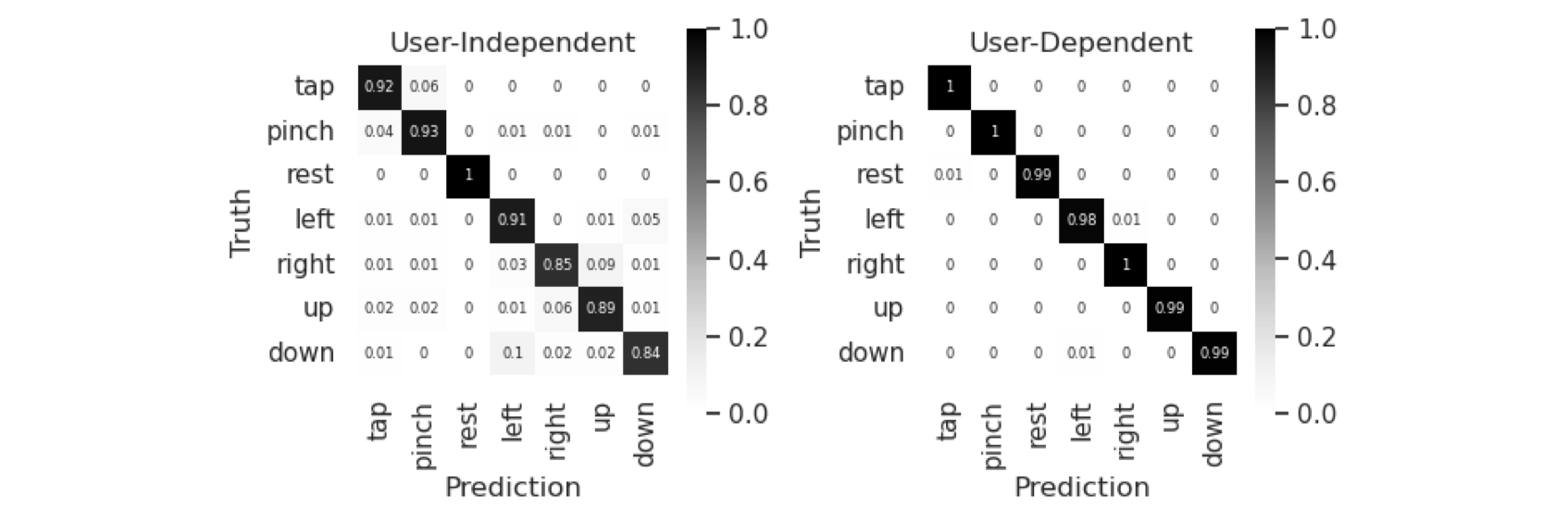}
  \caption{Micro-Finger Gesture Recognition Study Results. For the 7-class classification task, the user-independent model achieves an accuracy of 90.60\%. The user-dependent model achieves an accuracy of 99.27\% with 9.3 min of training data.}
  \label{fig: microresults}
\end{figure*}

\subsection{Procedure}
We conducted a user study with 10 participants, in which 2 participants each participated in 1 of the 2 previous studies.
The study lasted about 60 minutes with a similar procedure as the third study.
Each participant performed 10 sessions of gestures, and the first session was discarded as a practice session. Before each session, the participants remounted the ring by themselves.
From each participant, we collected: $7 \text{ gestures } \times 10 \text{ repetitions } \times 9 \text{ sessions } = 630 \text{ instances}$.
The participants were seated throughout the study with their hands naturally placed on the table.

\subsection{Results}
Similar to that in the previous study, we use LOPO cross-validation (with 3.15h of training data) to evaluate the user-independent performance for the 7-class classification.
Across 10 participants,~\theDevice{} has a mean accuracy of 90.60\% (confusion matrix depicted in left in Fig.~\ref{fig: microresults}).
We then fine-tune the user-independent models with the participant's data.
The mean accuracy quickly increases to 96.6\% with 2.3 min of training data and 99.27\% with 9.3 min of training data (confusion matrix depicted in right in Fig.~\ref{fig: microresults}).
We further compare the results with prior ring systems that recognize thumb-index micro-gestures in Table~\ref{table:micro-results}.
The promising results demonstrate that ~\theDevice{} detects fine-grain movements effectively. 
\begin{table}[t]
    \centering
    \caption{Thumb-to-index micro-gesture recognition performance comparison. ``UD'' stands for user dependence and ``UI'' stands for user independence.}
    \begin{tabular}{|c|c|S[table-format=2.2, table-column-width=3cm]|S[table-format=2.2, table-column-width=1.5cm]S[table-format=2.2, table-column-width=1.5cm]|}
    \hline
        System & Gesture Set & {UD Accuracy} & \multicolumn{2}{c|}{UI Accuracy} \\
    \hline
        Thumb-in-Motion, 2018~\cite{Thumb-In-Motion} & 5 gestures  & \textemdash &  & 89\%  F1\\
    \hline
        Z-ring, 2023~\cite{z-ring} & 9 gestures & 88\%  & 83.67\% &\\
    \hline
        EF-ring, 2023~\cite{efring} & 9 gestures & 89.5\% & 85.2\% &\\
    \hline
        \textbf{\theDevice{}} & 7 gestures & 99.27\%  & 90.60\% & 90.58\% F1 \\
    \hline
    \end{tabular}
    \label{table:micro-results}
\end{table}
\section{Discussion, Limitations and Future Work}\label{sec: discussion}
The user studies showcase that our system tracks hand positions continuously with resilience to varying noise factors and wrist orientations.
The micro-gesture recognition study and the input to AR glasses demo in the video figure serve as example uses of the pose tracking capabilities for practical one-hand interaction.
In the future, as the ring's tracking matures, more applications (\textit{e.g.}, sign language translation) will be unlocked.
As~\theDevice{} is the first single-ring system that enables hand pose tracking across sessions and users, it still has shortcomings that must be resolved before large-scale deployment in real-world settings.
For example, Gheran~\etal{}'s attempt to replicate a ring-based gesture elicitation study revealed~\cite{ring-gesture-elicitation} real-world replicability issues.
In this section, we discuss the limitations of ~\theDevice{} and the challenges and opportunities of broader smart ring systems.

\subsection{Real-Time Tracking Delay \& Performance.}
Our current system is deployed on a smartphone for real-time inference. However, as the demo video shows, we observe a noticeable delay. 
Though we already alleviated the delay by using the last frame in the sliding window as ground truth, the transmission and computing times are not negligible.
We measured the averaged delay using (a) the ring that captures and sends the acoustic signals to a smartphone via Bluetooth; (b) a smartphone (Redmi Note 12 Pro) that receives the signals, decodes the echo profiles, makes inferences with PyTorch Mobile~\footnote{https://pytorch.org/mobile/home/}, and sends the prediction to a computer via WiFi; and (c) a computer (Apple Macbook Air, 2022) that receives and processes the predictions for visualization or control purposes.
\begin{table}[t]
    \centering
    \caption{Real-time Tracking Delay Time Breakdown.}
    \begin{tabular}{|c|c|c|c|c|c|}
         \hline
         Step & BLE transmission& Echo Profile Calculation&  Inference&  WiFi Transmission& Hand Visualization \\
         \hline
         Time (ms)& 200.0 & 14.7 & 54.6-70.4 & 16.7 & 126.9-190.5 \\
         \hline
    \end{tabular}
    \label{table: delay}
\end{table}

We report in Table~\ref{table: delay} the time for each step.
For hand pose tracking, we found a total delay time of 413-492 ms depending on the processing availability of the phone and the computer, in which 126.9-190.5 ms attributed to the hand visualization.
The visualization time is long because it uses an inverse kinematic solver and renders a high-fidelity hand in 3-dimensions.
The visualization time can be omitted if the joint coordinates are directly used.
While the echo-profile calculation and inference times, 69.3-85.1 ms, are short, the BLE transmission takes 200 ms.
The selected Android phone has high Bluetooth latency as a result of the tradeoff with reliable high bandwidth.
In the future, this delay time can be further reduced with direct hardware control (e.g. nRF52840 Dongle) or advanced BLE hardware/protocols, or compressing data for a much smaller transmission package.

We also observed joint distortions and inconsistent internal torques in the predictions of \theDevice, stemming from the model lacking information about kinematic constraints.
A possible solution involves regressing pose parameters (e.g. MANO~\cite{mano}) as ground truth, that account for such kinematic constraints.
From our preliminary exploration of within-user models, the MANO representation exhibits an angular error of 11.36\textdegree, 0.6\textdegree smaller than the original MediaPipe representation. We do not compare the distance errors here as they vary in scale.
Future research in wearable hand-pose tracking could consider similar approaches or investigate alternative methods for acquiring high-fidelity ground truths, such as multi-camera systems like Quest~\cite{quest}.
\subsection{Model, Model Input, and Data Augmentation Selection.}\label{discussion: comparative}
Using the user-independent and user-dependent evaluation data collected and similar training schemes in the second user study, we compare our current approach with the alternatives.

\begin{figure}[h]
  \includegraphics[width=\textwidth]{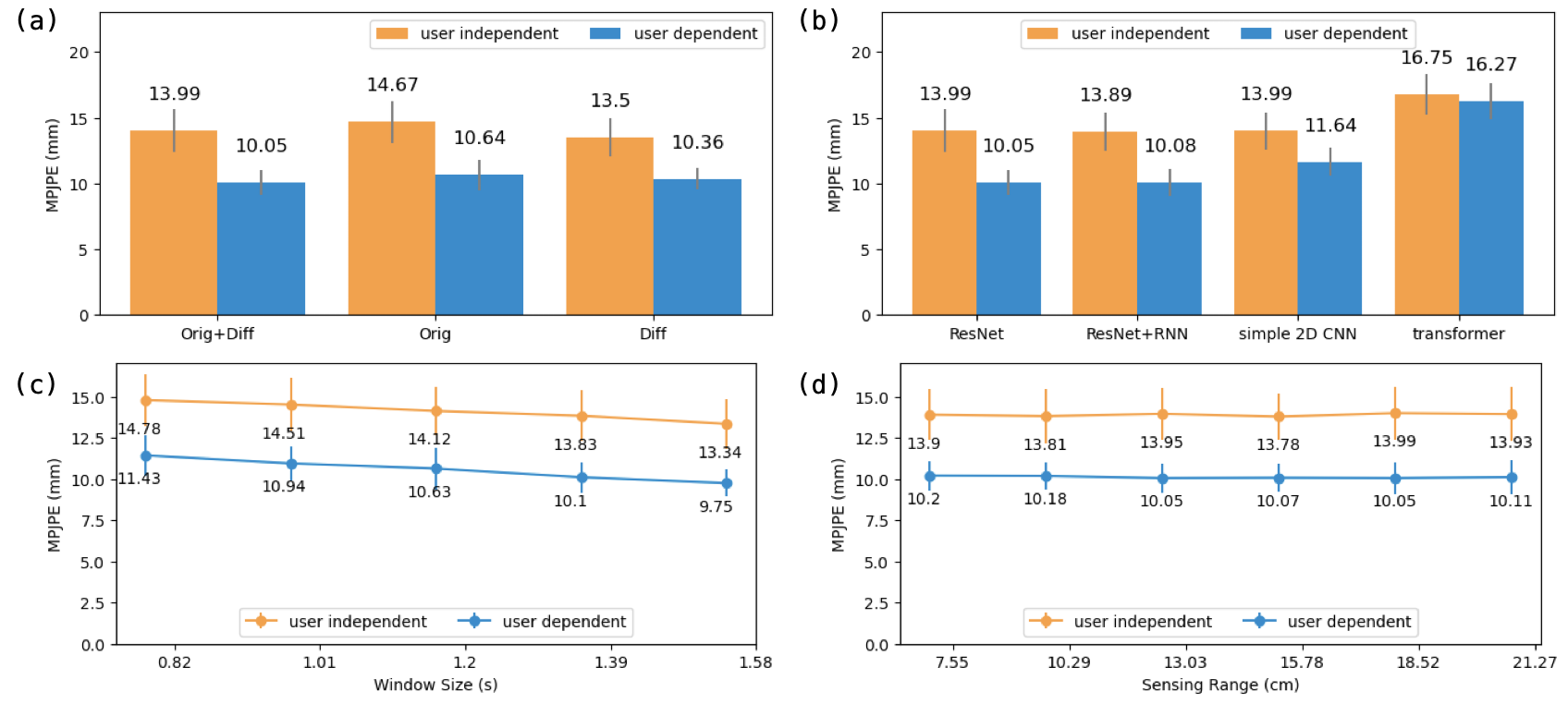}
  \caption{Comparative studies on the deep learning model and model inputs. (a): Echo profile ablation study: ``Orig'' denotes the original echo profile and ``Diff'' denotes the differential echo profile. (b): Model comparison study: Resnet, our current approach, and ResNet+RNN have better performances than simple 2D CNN and transformer. (c) Sliding window length comparison study: a reasonable increase in the window size increases the performance. (d) Sensing range comparison study: sensing range has little impact on the sensing performance. }
  \label{fig: comparative}
\end{figure}

\subsubsection{Model Comparison Study}
In addition to the adopted ResNet backbone model, we experimented with (a) more complex models, such as adding RNN (GRU/LSTM) layers after the ResNet encoder and replacing ResNet with a transformer encoder; and (b) simpler models, like a 3-layer CNN.
We show the results in Fig.~\ref{discussion: comparative}(a): ResNet (13.99mm for user-independent, 10.05mm for user-dependent) and ResNet+RNN (13.89mm, 10.08mm) similar performance.
The simple 2D CNN has equally good user-independent performance (13.99mm) but much worse user-dependent performance (11.64mm).
The transformer model has the worst performance for both independent (16.75mm) and dependent(16.17mm) evaluations.
We conjecture ResNet's good performance is a result of echo profiles' efficient 2D representation of time and distance, so the additional temporal learning is unnecessary.
ResNet was picked for the smallest errors and the lower computational costs, compared with ResNet+RNN.

\subsubsection{Echo Profile Ablation Study}
We used the combination of original and differential echo profiles to capture absolute hand shapes and movements.
In this ablation study (Fig.~\ref{fig: comparative}(b)), we train and test with only the original or differential echo profiles.
The two-way repeated ANOVA test with model type and input types as independent variables further shows statistical significance (F$_{2,22}$=65.0, p<0.001).
For user-independent models, only the original echo profile (14.67mm) has a larger error than the original+differential echo profiles (13.99mm), but the differential echo profile itself has even smaller errors.
This is attributed to the relatively large individual hand shape differences, in comparison to finger movements.
However, for the user-dependent model, both the original echo profile (10.64mm) and the differential echo profile (10.36mm) contribute heavily to the prediction.
This is because the additional fine-tuned data contains hand shape and movement information tailored to the user.

\subsubsection{Window Size Comparison Study} 
In our current system implementation, we chose a window length of 1.2 seconds.
However, the window size can potentially impact the performance.
The experiment results on different window sizes are shown in Fig.~\ref{fig: comparative}(c), his showed the performance improved as the window size increased: the user-independent and dependent errors decreased from 14.78mm and 11.43mm to 13.34mm and 9.75mm when the window size increased from 0.82s to 1.58s. The two-way repeated ANOVA test with model type and window size as independent variables further shows statistical significance (F$_{4,44}$=9.2, p<0.001).  

\subsubsection{Sensing Range Comparison Study}
The selection of sensing range affects the balance between captured noise and information. 
In our experiment comparing sensing ranges (Fig.~\ref{fig: comparative}(d)), we observed minimal impact (F$_{5,55}$=0.79, p=0.56 in two-way repeated ANOVA test with model type and sensing range as independent variables) on performance based on varying sensing ranges: errors of user-independent models range from 13.78mm to 13.99mm, and errors for user-dependent models range from 10.05 and 10.20.
Our choice of sensing range was determined by the average hand size. 
\begin{table}[t]
    \caption{Data Augmentation and Input Normalization Ablation Study Results. ``UD'' stands for user dependence, and ``UI'' stands for user independence. All errors refer to mean per-joint position error (MPJPE).}
    \centering
    \begin{tabular}{|c|>{\centering\arraybackslash}m{2cm}|>{\centering\arraybackslash}m{2cm}|>{\centering\arraybackslash}m{2cm}|>{\centering\arraybackslash}m{2cm}|}
        \hline
         & \begin{tabular}[c]{@{}c@{}}UI \\ Training\end{tabular}  &  \begin{tabular}[c]{@{}c@{}}UD \\ Training\end{tabular}& \begin{tabular}[c]{@{}c@{}}UI \\ Robustness\end{tabular} & \begin{tabular}[c]{@{}c@{}}UD \\ Robustness\end{tabular} \\
         \hline
         w/o data augmentation& 14.16 & 10.67 & 15.24 & 12.92 \\
         \hline
         w/o vertical shifts&  14.37 & 10.62 & 15.25 & 12.94 \\
         \hline
         w/o randomness& 14.07 & 10.48 & 14.76 &  12.68\\
         \hline
         w/o input norm& 14.34 & 10.43 & 15.06 & 12.63\\
        \hline
         \begin{tabular}[c]{@{}c@{}}w/ data augmentation\\ w/ input norm\end{tabular}&  14.12& 10.44 & 14.74 & 12.70\\
         \hline
    \end{tabular}
    \label{tab:aug-norm}
\end{table}

\subsubsection{Input Normalization Ablation Study}
As described in Sec.~\ref{sec:model-framework}, the original and differential echo profiles are each normalized to account for inconsistent signal magnitudes.
Using the same evaluation scheme as the main pose tracking study, we compare the user-independent and -dependent performance when tested with data collected in the same setting as the training data (UI and UD Training in Table~\ref{tab:aug-norm}) and when tested with data collected from the robustness testing sessions (UI and UD Robustness in Table~\ref{tab:aug-norm}).
For user-dependent models, input normalization has little effect.
Input normalization has larger improvements when signals vary more: 15.06$\rightarrow$14.74 for UI robustness with variations across users and settings, 14.34$\rightarrow$14.12 for UI testing with variations across users.

\subsubsection{Data Augmentation Study}\label{sec:aug-study}
For the pose-tracking task, we apply 2 data augmentation techniques: vertical shifts and randomness, as described in Sec.~\ref{sec:model-framework}.
We conduct the ablation study in the same manner as the above section by removing individual augmentation techniques and both augmentation techniques.
Without any data augmentation, the performance drops for all testing scenarios, proving effectiveness.
Similar to the observation in the input normalization ablation study, UI robustness has the largest improvement as it has the largest variations: 15.24$\rightarrow$14.74.
Introducing randomness on its own has little impact on the performance while applying vertical shifts consistently decreases the errors.
Future work on systematically investigating different data augmentation techniques has the potential to improve tracking abilities and increase robustness.

\subsection{Future Evaluation in Unconstrained Settings.}
Our current hand pose tracking study, capturing a wide range of noise factors, evaluates the system in a relatively controlled setting to demonstrate the feasibility of the proposed sensing system. 
Our pose set of size 20 is relatively large when compared among literature (\textit{e.g.}, 11 in ~\cite{etherpose} and 16 in~\cite{opisthenar}), but can not exhaust all possible human hand poses as a research prototype concept.
Though we included two uncontrolled hand movement sessions, one with a surface in the front, and one without, in the robustness testing section, the participants kept their hands within the camera view for ground truth acquisition, so the movement does not exactly replicate those in everyday activities.
For example, fast global hand movements in walking/running and finger movements may introduce a larger Doppler effect.
Our system does not directly analyze the frequency shift caused by the Doppler effect for speed or movement analysis.
Instead, our system learns the holistic profile of echoes for pose estimation. 
A limitation of this approach is that if the distribution of the training data significantly differs from that of the testing data, performance may decline. 
To address this issue, as with other data-driven approaches, we need to collect additional training data in these specific scenarios to augment the dataset.

We did not constrain the transition speeds in our user studies, but we did not formally evaluate the impact of transition speeds between hand poses with a hand motion speed study~\cite{EchoWrist}.
Although our evaluation protocol is similar or comparable to prior work~\cite{EchoWrist, etherpose}it remains important for future work to evaluate the system's performance in a true freestyle manner in the wild.

Another limitation is that we conducted the user study with adults only, but children and infants have smaller and thinner hands which introduces even more hand geometry variations. It is unclear how our system works with children and infants, beyond the scope of the paper as a proof-of-concept. 

Furthermore, in terms of location, our indoor studies do not account for outdoor settings where there could be ultrasonic interference.
Because ~\theDevice{} leverages reflection strengths information from within 18.52 cm, objects and other body parts (\textit{e.g.}, legs and hands) within the range alter the signals, as shown in the object robustness study.
This is a limitation shared by many existing wearable hand pose sensing systems.
Conversely, this ``noise'' might be seen as information regarding adjacent objects~\cite{EchoWrist} and the hand status. For instance, when an object is detected and moves closer to the hand, the reflected acoustic signal could possess a unique pattern that not only indicates the object's shape but also provides insights into the hand pose. It would be interesting to see other sensing tasks the ring can accomplish, such as object detection and activity recognition.

\subsection{Integration into Smart Ring Platforms.}

\paragraph{Energy Consumption.}
From the hardware perspective, the simplicity of our sensor setup allows easy integration of the hand pose tracking solution into an existing smart ring.
In Fig.~\ref{fig: hardware}(c), we show a modification of our prototype with an arc battery (Gepow GRP1507028) and without the microcontroller unit.
For instance, to integrate our solution into a commercial product like the Oura Ring~\cite{OuraRing}, which already features a speaker and Bluetooth capability, the addition of just a microphone would be enough.
However, there are other obstacles.
Though relatively low-power for a wearable hand-pose tracking solution (Table.~\ref{table:rings}), active acoustic sensing, while effective, does not present the absolute best energy efficiency for a ring.
The current power consumption of the device still only lasts about 1.75h with our 70mAh Lipo battery due to the (relatively) high energy consumption of our selected flat speaker. 
This speaker was a tradeoff between a slim form factor and a lower efficiency (we experimented with thicker speakers with much lower energy consumption).
The speaker's transmission power could also be lowered to reduce energy consumption.
The speaker's current transmitted signal reaches far enough to capture the whole hand's reflection, and our cropped sensing range is much smaller than the theoretical sensing range as described in Sec.~\ref{sec: sensing-principle}.
Our study on robustness to nearby objects further validates that when the object is within the sensing range in front of of the ring, the reflection indeed affects the signals.
Future works on alternative hardware and reduced speaker transmission power will make the device more energy efficient.

\paragraph{Multimodal Sensing.}
In this paper, we solely employed active acoustic sensing.
When integrated into a smart ring platform like ~\cite{OuraRing, onering}, multimodal sensing, using existing inertial measurement units (IMUs) and Photoplethysmography (PPG), can further enrich~\theDevice{}'s hand sensing capabilities~\cite{AppleWatch, onering}.
Other sensing principles will surely complement some of~\theDevice{}'s limitations: electric field sensing may correct the blockage interference within the ring's sensing range~\cite{efring}; IMUs/capacitive proximity sensor can provide additional movement information on the instrumented finger~\cite{learning-on-the-rings} and on additional two neighboring fingers~\cite{PeriSense}.
Additional sensors bring additional energy consumption and space constraints, posing challenges and opportunities for low-power and miniaturized sensor solutions.

\paragraph{Wearable Ecosystem.}
As we revisit the research question stated in the introduction, a single untethered ring succeeds in continuously tracking hand poses.
However, as an accessory, it is not uncommon for one to wear multiple rings on one hand and even on one finger.
When designing gestures for most prior single-ring systems~\cite{efring, thumbtrak, ElectroRing21}, the worn finger is heavily involved in performing the gestures for accurate sensing results, but our micro-gesture study demonstrated that a single ring effectively distinguishes gestures performed by the other fingers, enabling a larger design space for smart ring gestures with fewer constraints.
Current multi-ring systems~\cite{learning-on-the-rings, TelemetRing} contain a wristband in addition to multiple rings.
Building on top of~\theDevice{}, it would be interesting to explore the design, technical, and interaction space of multiple rings where users choose their number of rings and placements.
Further, we envision that in the future, the ring will work together with other wearables as part of the ecosystem to provide users with the optimal always-available interaction experience.

\section{Conclusion}
In this paper, we present, ~\theDevice{}, the first smart ring that tracks hand poses continuously and recognizes hand postures using inaudible active acoustic sensing. A series of 3 user studies with 36 participants showed that it can achieve ~14.1 mm accuracy in inferring the relative (to the wrist) positions of 20 finger joints without the need to collect training data from a new user. The joint error decreases to 10.3 mm with additional calibration data from the user. The promising results pave the way for ring-based hand pose sensing systems. 

\begin{acks}
This project was supported by the National Science Foundation Grant No. 2239569, and partially by the National Science Foundation’s I-Corps Award No.
2346817 and the Cornell University IGNITE Innovation Acceleration Program.
We want to thank the study
participants and the reviewers.
ChatGPT was utilized to polish the paper writing.
\end{acks}
\bibliographystyle{ACM-Reference-Format}
\bibliography{ring}

\end{document}